\documentclass[]{aa} 
 
\usepackage{graphicx, times, amsfonts}

\newcommand{\rsol}{\mbox{R$_{\odot}$}}

\newcommand{\ks}{km s$^{-1}$}

\begin{document} 
 
 
\title{
Baade-Wesselink distances and the effect of metallicity in classical cepheids
}  
 
\author{ 
M.A.T.~Groenewegen 
\inst{1,2}  
}

\institute{ 
Koninklijke Sterrenwacht van Belgi\"e, Ringlaan 3, B--1180 Brussels, Belgium \\ \email{marting@oma.be}
\and
Instituut voor Sterrenkunde, Celestijnenlaan 200 D,  
B--3001 Leuven, Belgium 
} 
 
\date{received: 2008,  accepted:  2008} 
 
\offprints{Martin Groenewegen} 
 
\authorrunning{M.~Groenewegen} 
\titlerunning{Baade-Wesselink distances and the effect of metallicity in cepheids} 
 
\abstract{The metallicity dependence of the Cepheid $PL$-relation is
  of importance in establishing the extra-galactic distance scale.}
{The aim of this paper is to investigate the metallicity dependence of
  the $PL$-relation in $V$ and $K$ based on a sample of 68 Galactic
  Cepheids with individual Baade-Wesselink distances (some of the
  stars also have an HST-based parallax) and individually determined
  metallicities from high-resolution spectroscopy.}
{Literature values of the $V$-band, $K$-band and radial velocity data
  have been collected for a sample of 68 classical cepheids that have
  their metallicity determined in the literature from high-resolution
  spectroscopy. Based on a $(V-K)$ surface-brightness relation and a
  projection factor derived in a previous paper, distances have been
  derived from a Baade-Wesselink analysis. $PL$- and $PLZ$-relations
  in $V$ and $K$ are derived.}
{The effect of the adopted dependence of the projection factor on
  period is investigated. The change from a constant $p$-factor to one
  recently suggested in the literature with a mild dependence on $\log P$ 
  results in a less steep slope by 0.1 unit, which is about the
  1-sigma error bar in the slope itself. The observed slope in the
  $PL$-relation in $V$ in the LMC agrees with both hypotheses. In $K$
  the difference between the Galactic and LMC slope is larger and
  would favour a mild period dependence of the $p$-factor. The
  dependence on metallicity in $V$ and $K$ is found to be marginal,
  and independent of the choice of $p$-factor on period. This result
  is severely limited by the small range in metallicity covered by the
  Galactic Cepheids. }
{}


\maketitle

\section{Introduction} 
 
Obtaining accurate distances to stars is a non-trivial matter. 
Cepheids are considered an important standard candle as they are
bright and are thus the link between the distance scale in the nearby
universe and that further out via those galaxies that contain both Cepheids and SNIa.

Distances to local cepheids may be obtained via main-sequence fitting
for those Cepheids in clusters (e.g. Feast 1999) or via determination
of the parallax.  Until recently only Polaris had an accurate parallax
determination via Hipparcos.  Benedict et al. (2007) recently
published absolute trigonometric parallaxes for nine Galactic Cepheids
using the {\it Fine Guidance Sensor} on board the {\it Hubble Space
  Telescope}, and revised, more accurate, Hipparcos parallaxes have
also become available (van Leeuwen et al. 2007, van Leeuwen 2007).

In addition, distances to Cepheids can be obtained from the
Baade-Wesselink method. This method relies on the availability of
surface-brightness (SB) relations to link variations in colour to variations
in angular diameters and understanding of the projection ($p$-)
factor that links radial velocity  to pulsational velocity variations.

The SB relations can be obtained using Cepheids that have accurate
interferometrically determined angular diameters over the pulsation
phase and accurate multi-wavelength light curves.  This has allowed
Kervella et al. (2004) and Groenewegen (2007; hereafter G07) to derive
such relations, in particular using the $(V-K)$ colour, which gives the
highest precision in the derived angular diameter.

The $p$-factor can also be obtained from Cepheids with
interferometrically determined angular diameters over the pulsation
phase when in addition accurate RV curves are available (see
e.g. M\'erand et al. 2005). G07 assembled all currently available data
to find that a constant $p$-factor represents the best fit to the
available data. A strong dependence on period, like $p \sim - 0.15
\log P$ as proposed by Gieren et al. (2005), could be ruled out. A
moderate dependence of $p \sim - 0.03 \log P$ as used in the Gieren et
al. papers (1993, 1997, 1998, Storm et al. 2004, Barnes et al. 2003),
or $p \sim - 0.064 \log P$ as advocated by Nardetto et al. (2007) are
also consistent with the available data.

Another aspect of the Cepheid $PL$-relation that is still a matter of
debate is the metallicity dependence of the Cepheid $PL$-relation, and
its dependence on wavelength.
Observations seem to consistently indicate that metal-rich Cepheids
are brighter, and various estimates have been given in the literature,
$-0.88 \pm 0.16$ mag/dex ($BRI$ bands, Gould 1994),
$-0.44^{+0.1}_{-0.2}$ mag/dex ($VR$ bands, Sasselov et al. 1997),
$-0.24 \pm 0.16$ mag/dex ($VI$ bands, Kochanek 1997), 
$-0.14 \pm 0.14$ mag/dex ($VI$ bands, Kennicutt et al. 1998), 
$-0.21 \pm 0.19$ in $V$, $-0.29 \pm 0.19$ in $W$, $-0.23 \pm 0.19$ in $I$, 
$-0.21 \pm 0.19$ mag/dex in $K$ (Storm et al. 2004), and
$-0.29 \pm 0.10$ mag/dex ($BVI$ bands, Macri et al. 2006).

The potential caveat is that no individual abundance determinations of
individual Cepheids are being used in these studies but rather
abundances of nearby H{\sc ii} regions, or even a mean abundance of
the entire galaxy.

Groenewegen et al. (2004) tried to attack the problem by using
Cepheids with individually determined metallicities, 37 in our Galaxy,
10 in the LMC and 6 in the SMC. Considering the Galactic sample, a
metallicity effect was found in the zero point in the $VIWK$
$PL$-relation of $-0.6 \pm 0.4$ or $-0.8 \pm 0.3$ mag/dex, depending on
the in- or exclusion of one object. The distances to the galactic
Cepheids mostly came from BW-distances available at that time from
Fouqu\'e et al. (2003).

In the last few years significant progress has been made in several areas. 
On the one hand more Cepheids have interferometrically determined
angular diameters. Combined with improved theoretical studies
(Nardetto et al. 2007, and references therein) this has led to an improved
understanding of the $p$-factor and SB relations, as was studied in G07.
On the other hand, more basic photometric light curves and radial
velocity data have become available and new studies have presented
metallicity determinations for significant samples (see references below).

In this paper, Baade-Wesselink distances to Cepheids with metallicity
determinations are revisited, using the $p$-factor and the SB $(V-K)$
colour relation derived in G07, updated as described in Sect.~3.
Section~2 describes the selection of the photometric and radial
velocity data.  Section~3 outlines how the data are modelled. 
Section~4 describes how the binary Cepheids are treated, and new and
updated orbital elements are presented. Section~5 describes the results. 
Section~6 presents conclusions.

\section{The sample} 
 
In a first step Cepheids with recent (post-2000) accurate metallicity
determinations were selected (from 
Andrievsky et al. 2002abc, 2003, 2004, 2005,
Luck et al. 2003, 2006, 
Luck \& Andrievsky 2004, 
Kovtyukh et al. 2005a,b, 
Yong et al. 2006,  Lemasle et al. 2007, Mottini et al. 2008), and for 
those, the McMaster Cepheid Photometry and Radial Velocity Data
Archive\footnote{http://crocus.physics.mcmaster.ca/Cepheid/} was
checked for the availability of $V$, $K$ and radial velocity (RV) data. 
For the stars selected, the literature was searched for additional data.

Table~\ref{TAB-DATA} lists the 68 stars selected with the source of
data indicated.  The last column indicates if particular datasets
and/or ranges in Julian date (JD) were not used in the fitting
described in the next section to derive the distance. This is
typically because these datasets show very large residuals, or the
stars show a noticeable change or shift in period, and the range in JD
had to be restricted, typically centred on the available $K$-magnitude
data. For the determination of the orbital parameters of the Cepheids
in binaries (Sect~\ref{sect-bin}), all available RV data were used.

For four Cepheids new high-resolution spectra were obtained in order
to derive the radial velocity\footnote{Two stars are no longer considered
  in the Baade-Wesselink analyses below: 
BQ Ser and V367 Sct which 
are classified as double-mode Cepheids.}. Observations took place in July and
August 2007 using the Coralie spectrograph at the 1.2m Euler telescope
located at the La Silla observatory, Chile. This is a fibre-fed
echelle spectrograph (2\arcsec\ fibres with one on the object and one
on the sky), which covers the 3880 to 6810\AA\ region in 68 orders
with a spectral resolution of about 50~000. Exposure times are between
3 and 10 minutes.  During each night, several exposures with a
tungsten lamp were taken to measure the relative pixel sensitivity
variation of the CCD. A Thorium-Argon lamp was observed at the
beginning and end of the night for wavelength calibration.
The Heliocentric velocities are obtained as part of the standard
pipeline reduction whereby the spectrum is cross-correlated against a
mask, in this case that of a G2 star. The formal precision on the RVs
is of the order 5-15 m/s. The RVs are listed in Table~\ref{Tab-Dat}.

Like in G07, in a first step to attempt to homogenise the
datasets, the light curves are read in the program Period04 (Lenz \&
Breger 2005), which allows for an easy visualisation of different
datasets. Period04 is used to calculate the period, and phase the data to
see if there are clear outliers in the data (which are flagged and
excluded) or clear off-sets between datasets.
Some datasets were in the end not considered because of very large
error bars. The rms in the fit was determined for each dataset
separately and this was used as the typical error bar (unless the rms
was consistent with the original error bars quoted, in which case these
were retained).

$K$-band data on the CIT/CTIO system were transformed to the
SAAO-Carter system using the formula in Carter (1990), and Johnson and
SAAO-Glass IR-photometry was transformed to the SAAO-Carter system
according to Glass (1985).

Table~\ref{Tab-metal} lists the adopted metallicity value. 
They were  taken preferentially from Luck \& Andrievsky (2004),
Kovtyukh et al. (2005), and Andrievsky et al. (2005) who studied the
fundamental parameters, including abundances, at several phases.
Alternatively, values were taken from other papers by
Andrievsky/Luck/Kovtyukh, and averaged when multiple determinations
were available.
Thirdly, iron abundances from Fry \& Carney (1997) and Yong et al. (2006) 
were transformed to the ``Andrievsky system'' following 
Laney \& Caldwell (2007) and Luck et al. (2006).  Based on the 12
stars in common between Andrievsky et al. (2003) and Mottini et al. (2008) 
an offset AND - MOT = +0.06 $\pm$ 0.11 was determined, and based on 6
stars in common with Lemasle et al. (2007) an offset AND - LEM = +0.09
$\pm$ 0.07.  These offsets were applied to the Mottini et al. and
Lemasle et al. iron abundances to bring them to the ``Andrievsky system''.




\begin{table*} 

\caption{Sources of $V$-, $K$-band and RV data. } 
\begin{tabular}{rrrrl} \hline \hline 

Name       & $V$                               & $K$            & RV            &  Data not considered$^1$ \\  
\hline 
AQ Pup     & 1,2,3,4,5,84 & 6,7,8   & 1,10,11,44     & 3,-,- JD $<$ 34000, JD $>$ 49700 \\
BB Sgr     & 2,3,16,63    & 6,24    & 20,33,39,44    & -,-,44 \\
beta Dor   & 3,11,16      & 6,18    & 11,19,20,21,22 & -,-,20     \\  
BF Oph     & 2,3,16,23,63 & 6,18,24 & 10,20,39,44,76 & JD $<$ 36000 \\
BG Lac     & 2,3,12,13    & 12      & 14,15,44 \\
BN Pup     & 1,3,5,16     & 6,7     & 1,17 \\                                                  
CD Cyg     & 2,3,64       & 24      & 33,44 \\
CV Mon     & 2,3,13,16,23    & 6,24  & 25,26,44 \\
delta Cep  & 2,3,11,12,13,27 & 12,98    & 4,9,14,25,28,29 & -,-,9 JD $<$ 43500 \\              
DT Cyg     & 2,3,4,27,30,31  & 24,32 & 4,9,28,29,33,34 & -,32,9+28 JD $<$ 34000 \\             
eta Aql    & 2,3,12,13,16,27,35 & 12,24 & 9,14,20,25,28,29,36,37 & -,-,9 \\
FF Aql     & 2,3,11,16,27,30,31 & 18,24,32 & 9,20,28,29,33,38 & 3,32,9 JD $<$ 44000 \\         
FM Aql     & 2,3,12,13,16,23 & 12,24 & 14,33,39,44 & JD $<$ 30000 \\
FN Aql     & 2,3,12,13,16,23 & 12,24 & 14,33,39    & -,-,39 \\
KN Cen     & 1,3,5,11,16,40,84 & 6 & 1,17,41 & -,-,41 JD $<$ 40000, JD $>$ 47000 \\            
KQ Sco     & 1,3,11            & 6 & 1,42 \\                                                   
l Car      & 3,5,11,16,35   & 6 & 11,19,20,23,43 & 5,-,23 JD $<$ 46500 \\ 
LS Pup     & 1,3,11         & 6 & 1 \\                                                         
RS Pup     & 2,3,5,11,16,84 & 6,8,24 & 11,19,25,44 & -,-,44 JD $<$ 42000, JD $>$ 50000 \\
RT Aur     & 2,12,27,31,45  & 12,32 & 9,28,29,33,37,46 & -,-,9  \\                             
RU Sct     & 2,3,16         & 6     & 26,33,105 &                   \\
RY Sco     & 1,2,3,5,11,16  & 6     & 1,20,44 \\
RY Vel     & 1,3,5,11,16    & 6     & 1,11,17 & JD $<$ 44000, JD $>$ 50000 \\
RZ Vel     & 1,3,5,11,16    & 6 & 1,11,20,47 \\
S Mus      & 3,11,23,40     & 6,24 & 10,11,20,22,41,47,48,49,50 & -,-,41 JD $<$ 30000 \\        
S Nor      & 3,11,23,40     & 6,24 & 4,10,20,41,47,49,51,52,53  & -,-,41 JD $<$ 30000 \\        
S Sge      & 2,3,11,12,13,23,27,30,40 & 12,24,32 & 9,14,28,29,33,37,54,55,56 & -,32,54 \\       
SS Sct     & 2,3,16,23,63   & 24 & 39,42,44,63 & JD $<$ 30000 \\                                
SU Cas     & 2,3,9,27,31    & 12,32 & 4,9,25,28,29,33,57,58,59,60,61,62 & -,32,9 JD $<$ 43000 \\  
SU Cyg     & 2,3,27,30,31   & 24    & 9,28,33,106,107,108,109           & -,-,9  JD $<$ 35000 \\  
S Vul      & 3,30           & 6     & 33, 110 & JD $<$ 45400 \\
SV Vul     & 2,3,4,9,27,64  & 6,12,24 & 4,9,14,15,25,28,29,33,39,65 & JD $<$ 45500, JD $>$ 48600 \\ 
SW Vel     & 1,3,5,11,16,84 & 6    & 1,11 \\
SX Vel     & 3,23,40        & 6    & 10   \\                                                    
SZ Aql     & 2,3,11,12,16,23,64 & 6,12,24 & 11,14,23 \\
SZ Tau     & 2,3,4,12,27,31 & 6,12,32 & 4,9,28,29,33 & -,32,9 JD $<$ 43000, JD $>$ 48000\\
T Mon      & 1,2,3,11,16,35,64  & 6,24,32 & 4,9,25,29,33,37,48,65,66,67,68,69,70,71,72,73 & -,32,70 JD $<$ 44500 \\ 
TT Aql     & 2,3,11,12,16,27,64,74 & 12,24 & 9,11,14,15,22,28,33,37,39,74 & -,-,74 JD $<$ 44000 \\
T Vel      & 3,16,75        & 6,18 & 10,20,75 & -,18,- \\
T Vul      & 2,3,12,27,31   & 12,24,32,98 & 4,9,14,28,29 & -,32,9 JD $<$ 43600\\
U Car      & 1,3,5,11,16,35 & 6,24 & 1,11,20 \\
U Nor      & 1,5,11,16      & 6    & 1,11,17,41 \\
U Sgr      & 2,3,4,16,23,35,63 & 6,24,32 & 4,9,10,11,20,25,28,33,39,47,51,76,77,78 & -,32,9 JD $<$ 37000 \\
UU Mus     & 1,5,11,40,84      & 6    & 1,11 \\
U Vul      & 2,3,12,13,23,27   & 12   & 4,14,33,37,44,79,80 & -,-,44 \\                        
\hline 
\end{tabular} 
\end{table*}

\setcounter{table}{0}
\begin{table*} 

\caption{Continued } 
\begin{tabular}{rrrrl} \hline \hline 

Name       & $V$                               & $K$            & RV         & Data not considered$^1$  \\  
\hline 
V340 Nor   & 3,4,82             & 6     & 4,26  \\                                             
V350 Sgr   & 2,3,23,63,99       & 18,24 & 20,22,33,44,76,83,39 & -,24,- JD $<$ 42000 \\        
V496 Aql   & 2,3,16,63          & 24    & 9,10,20,33,76 \\                                     
V Car      & 3,16,23            & 6     & 10.20,47 \\
V Cen      & 3,11,16,23,63      & 6,24  & 10,11,20,47,76 & -,24,- JD $<$ 42000 \\
VW Cen     & 1,3,16,84          & 6     & 1,41  & -,-,41 \\                                    
VY Car     & 1,3,5,11,16,99,100 & 6,24  & 1,11,20,41 & JD $<$ 42000 \\
VZ Cyg     & 2,3,12,31          & 12,24 & 4,14,33,39,44 & JD $<$ 40000 \\                      
VZ Pup     & 1,3,5,11,16        & 6,7,8 & 1,11 \\
W Sgr      & 2,3,40,85          & 24,32,86 & 4,12,20,22,28,47,48,85,87,88,89 \\                
WZ Car     & 1,3,5,11,16        & 6     & 1,11 \\
WZ Sgr     & 1,2,3,5,11,40      & 6     & 1,11,33,39 & JD $<$ 40000 \\
X Cyg      & 2,3,27             & 12,24,32,98 & 4,14,25,29,33 \\ 
X Lac      & 2,3,12,13          & 12          & 4,14,33,44 & -,-,44 \\
X Pup      & 2,3,5,11,16,23     & 6           & 11,23,44   & 5,-,- \\
X Sgr      & 2,3,11,16,35,99    & 24,32,104 & 9,10,11,20,28,47,52,101,102,103 & -,-,9 JD $<$ 35000 \\ 
XX Cen     & 3,11,40,74         & 6,24  & 10,11,20,41,52,74 & -,24,74 \\                       
Y Lac      & 2,3,12,31,90       & 12    & 14,15,33,39,44 &  JD $<$ 26500 \\
Y Oph      & 1,2,3,16           & 6,24  & 1,9,19,20,28,33,34,91,92 & -,-,9 JD $<$ 40000 \\
Y Sgr      & 2,3,11,16,23,35    & 24,32 & 9,11,19,20,28,46,49,93,94 & -,-,9 JD $<$ 28500 \\    
YZ Sgr     & 2,3,23,40          & 24    & 20,44 \\
Z Lac      & 2,3,12,64          & 12    & 14,33,39,44,80,95,96,97 & JD $<$ 40500  \\           
zeta Gem   & 2,3,4,27,35,64     & 32,104 & 4,9,19,28,29,33,37 & -,-,9 \\                       
\hline 
\end{tabular} 

References:

1= Coulson \& Caldwell (1985a);   
2= Moffett \& Barnes (1984); 
3= Berdnikov et al. (2000), a datafile named ``cepheids-16-03-2006'' was retrieved from the ftp address listed in that paper;
4= Bersier et al. (1994); 
5= Madore (1975); 
6= Laney \& Stobie (1992); 
7= Schechter et al. (1992);
8= Welch (1985);  
9= Barnes et al. (1987), points with uncertainty flag ``:'' were removed; 
10= Stibbs (1955); 
11= Bersier (2002), data points with weight 0 and 1 in the Geneva photometry were removed;  
12= Barnes et al. (1997);  
13= Szabados (1980);   
14= Barnes et al. (2005); 
15= Imbert (1999);
16= Pel (1976); 
17= Pont et al. (1994);
18= Lloyd Evans (1980a); 
19= Nardetto et al. (2006); 
20= Lloyd Evans (1980b);  
21= Taylor \& Booth (1998); 
22= Petterson et al. (2005); 
23= Caldwell et al. (2001);  
24= Welch et al. (1984);  
25= Storm et al. (2004); 
26= Metzger et al. (1992); 
27= Kiss (1998); 
28= Wilson et al. (1989); 
29= Kiss (2000); 
30= Szabados (1991);   
31= Szabados (1977);  
32= Wisniewski \& Johnson (1968); 
33= Gorynya et al. (1998, VizieR On-line Data Catalog: III/229); 
34= Sanford (1935); 
35= Shobbrook (1992); 
36= Jacobsen \& Wallerstein (1981); 
37= Evans (1976); 
38= Evans et al. (1990);  
39= Joy (1937); 
40= Walraven et al. (1964); 
41= Grayzeck (1978); 
42= This paper, Table~\ref{TAB-DATA};
43= Taylor et al. (1997); 
44= Barnes et al. (1988), points with uncertainty flag ``:'' were removed;  
45= Turner et al. (2007); 
46= Duncan (1908); 
47= Lloyd Evans (1968);  
48= Petterson et al. (2004); 
49= Campbell \& Moore (1928); 
50= B\"ohm-Vitense et al. (1990); 
51= Mermilliod et al. (1987);
52= Feast (1967); 
53= Breger (1970); 
54= Evans et al. (1993);
55= Herbig \& Moore (1952);
56= Breitfellner \& Gillet (1993);
57= Adams \& Shapley (1918); 
58= Abt (1959);
59= Niva \& Schmidt (1979);
60= Gieren (1976);
61= H\"aupl (1988);
62= Beavers \& Eitter (1986);
63= Gieren (1981b);
64= Szabados (1981);  
65= Sanford (1956);
66= Evans et al. (1999); 
67= Frost (1906);
68= Sanford (1927);
69= Wallerstein (1972); 
70= Coulson (1983); 
71= Evans \& Lyons (1994);
72= Gieren (1989); 
73= Harper (1934); 
74= Coulson et al. (1985); 
75= Gieren (1985); 
76= Gieren (1981a); 
77= Jacobsen (1970);
78= Breger (1967);
79= Sanford (1951);
80= Imbert (1996); 
81= Coulson \& Caldwell (1985b); 
82= Eggen (1983); 
83= Evans \& Sugars (1997); 
84= Hipparcos Epoch Photometry;
85= Babel et al. (1989); 
86= Kimeswenger et al. (2004); 
87= Albrow \& Cottrell (1996); 
88= Jacobsen et al. (1984); 
89= Jacobsen (1974); 
90= Henden (1980); 
91= Evans \& Lyons (1992); 
92= Abt \& Levy (1978); 
93= Duncan (1922);
94= ten Bruggencate (1930); 
95= Evans \& Welch (1993); 
96= Gieren (1989);
97= Sugars \& Evans (1996); 
98= Fernley et al. (1989);
99= Dean (1977); 
100= Dean (1981); 
101= Duncan (1932); 
102= Slipher (1904);
103= Moore (1909); 
104= Feast et al. (2008);
105= Metzger et al. (1993);
106= Evans (1988);
107= Maddrill (1906);
108= Abt (1973);
109= Imbert (1985);
110= Joy (1952).

Note: 1= The number indicates the dataset not used in, respectively, $V$, $K$, RV.

\label{TAB-DATA} 
\end{table*}

\begin{table} 

\caption{New radial velocity data } 
\begin{tabular}{crcr} \hline \hline 

JD          & RV  (\ks) & JD          & RV  (\ks)    \\  
\hline 

\multicolumn{2}{c}{KQ Sco}  & \multicolumn{2}{c}{V367 Sct} \\
2454302.678 &  -7.57  &  2454341.605 &  -22.61   \\
2454303.670 & -10.77  &  2454341.605 &   26.97   \\
2454304.527 & -26.89  &  2454302.737 &  -19.57   \\
2454307.627 & -45.12  &  2454303.709 &  -19.15   \\
2454308.672 & -45.51  &  2454305.682 &    0.77   \\
2454309.722 & -53.38  &  2454306.754 &    4.38   \\
2454310.570 & -49.50  &  2454307.780 &  -16.31   \\
2454311.711 & -48.46  &  2454308.741 &  -16.90   \\
2454312.605 & -50.97  &  2454310.747 &   -5.25   \\
2454332.497 & -12.06  &  2454311.733 &   -7.84   \\
2454336.489 & -47.39  &  2454312.762 &   -9.21   \\
2454339.524 & -42.66  &  2454332.588 &    5.21   \\
2454341.537 & -46.58  &  2454334.567 &  -20.65   \\
            &         &  2454334.581 &  -20.30   \\
\multicolumn{2}{c}{SS Sct}  & 2454339.626 &  -13.20   \\
2454302.580 &    7.55  & 2454341.580 &   -9.54   \\
2454303.739 &  -24.37  &             &           \\
2454304.754 &  -11.28  & \multicolumn{2}{c}{BQ Ser} \\
2454305.772 &    3.70  & 2454302.601 &  -18.97   \\
2454306.711 &    1.72  & 2454303.731 &  -11.41   \\
2454307.691 &  -23.22  & 2454304.724 &  -15.08   \\
2454308.624 &   -8.05  & 2454305.738 &  -11.01   \\
2454309.616 &    5.48  & 2454306.731 &   -1.57   \\
2454310.785 &  -17.19  & 2454307.747 &  -27.11   \\
2454311.656 &  -18.74  & 2454308.717 &  -17.00   \\
2454312.806 &   -0.36  & 2454309.737 &   -4.14   \\
2454331.699 &    6.05  & 2454310.705 &   -2.98   \\
2454332.556 &   -4.74  & 2454311.753 &  -22.50   \\
2454334.613 &   -3.38  & 2454312.718 &  -11.03   \\
2454335.533 &    7.01  & 2454331.653 &    1.03   \\
2454336.671 &  -23.15  & 2454332.517 &  -23.78   \\
2454339.644 &    5.00  & 2454334.536 &  -10.36   \\
2454341.637 &   -8.36  & 2454335.515 &  -13.03   \\
            &          & 2454339.557 &   -6.35   \\
            &          & 2454341.559 &  -25.73   \\

\hline 
\end{tabular} 
\label{Tab-Dat}
\end{table}

\begin{table} 
\caption{Adopted [Fe/H] values for the Cepheids in the sample} 
\begin{tabular}{rrrr} \hline \hline 
Name       & [ Fe/H]  & Name       & [ Fe/H] \\  
\hline 

AQ Pup   & -0.14 (5)  &  SZ Aql   & +0.17 $\pm$ 0.04 (2) \\
BB Sgr   & +0.08 (6)  & SZ Tau   & +0.08 (4) \\ 
beta Dor & -0.01 (4)  &  T Mon    & +0.15 $\pm$ 0.03 (2) \\
BF Oph   & +0.00 (6)  & TT Aql   & +0.10 $\pm$ 0.02 (2) \\
BG Lac   & -0.01 (4)  & T Vel    & -0.02 (5) \\ 
BN Pup   & +0.01 (4)  &  T Vul    & +0.01 $\pm$ 0.02 (3) \\
CD Cyg   & +0.11 $\pm$ 0.05 (2)  & U Car    & +0.23 (7) \\ 
CV Mon   & -0.03 (4)  &  U Nor    & +0.13 (7) \\ 
del Cep  & +0.08 (4,6)  &  U Sgr    & +0.08 $\pm$ 0.02 (1) \\
DT Cyg   & +0.11 (4)  & UU Mus   & +0.17 (7) \\ 
eta Aql  & +0.08 $\pm$ 0.02 (1)  & U Vul    & +0.11 $\pm$ 0.03 (1) \\
FF Aql   & +0.02 (4) & V340 Nor & +0.00 (4) \\ 
FM Aql   & +0.08 (4) & V350 Sgr & +0.18 (6) \\
FN Aql   & -0.02 (4)  & V496 Aql & +0.05 (4)\\
KN Cen   & +0.23 (7)  &  V Car    & +0.02 (7,8) \\ 
KQ Sco   & +0.16 (4)  &  V Cen    & +0.04 (4) \\  
l Car    & +0.13 (7,8)  &  VW Cen   & +0.04 (7) \\ 
LS Pup   & -0.10 (7)  &  VY Car   & +0.03 (8) \\ 
RS Pup   & +0.17 (5)  &  VZ Cyg   & +0.05 (6) \\
RT Aur   & +0.06 (4) & VZ Pup   & -0.15 (4) \\  
RU Sct   & +0.01 (4) & W Sgr    & +0.03 $\pm$ 0.03 (1) \\
RY Sco   & +0.09 (6)  & WZ Car   & +0.14 (7) \\ 
RY Vel   & -0.03 (5)  & WZ Sgr   & +0.20 $\pm$ 0.05 (2)\\
RZ Vel   & -0.07 (5)  &     X Cyg    & +0.10 $\pm$ 0.03 (2) \\
S Mus    & +0.19 (7)  &   X Lac    & -0.02 (6) \\
S Nor    & +0.05 (4)  &   X Pup    & -0.03 $\pm$ 0.03 (2) \\
S Sge    & +0.10 $\pm$ 0.02 (1) & X Sgr    & -0.29 (4)\\
SS Sct   & +0.06 (6)  & XX Cen   & +0.10 (7) \\ 
SU Cas   & -0.01 (4)  & Y Lac    & -0.04 $\pm$ 0.06 (3)\\
SU Cyg   & -0.03 $\pm$ 0.04 (3)  & Y Oph    & +0.05 (4) \\
S Vul    & -0.03 $\pm$ 0.03 (2)  & Y Sgr    & +0.05 $\pm$ 0.03 (3)\\
SV Vul   & +0.05 $\pm$ 0.04 (2)  & YZ Sgr   & +0.08 (1) \\
SW Vel   & -0.03 (4,5)  &  zeta Gem & +0.04 (4) \\ 
SX Vel   & -0.03 (5)  &   Z Lac    & +0.01 $\pm$ 0.03 (2) \\

\hline 
\end{tabular} 

1= Luck \& Andrievsky (2004);
2= Kovtyukh et al. (2005b);
3= Andrievsky et al. (2005);
4= Andrievsky et al. (2003); 
5= Luck et al (2003); 
6= Luck et al. (2006); 
7= Mottini et al. (2008);
8= Lemasle et al. (2007).

\label{Tab-metal}
\end{table}

\section{The model}

The model is outlined in G07 but will be briefly repeated here. 
The $V$-, $K$- and RV data with error bars are fitted with a function of the form:
\begin{equation}
F(t) = F_0 + \sum_{i=1}^{i=N} 
\left( A_{\rm i} \sin (2 \pi \; t \; e^{i f}) + 
       B_{\rm i} \cos (2 \pi \; t \; e^{i f}) \right)
\end{equation}
where $P = e^{-f}$ is the period (in days). Typically, $P$ is
determined from the fit to the available optical photometry as this
dataset is usually most extensive. The period is then fixed when
fitting Eq.~1 to the $K$-band and RV data.
 
The determination of the parameters is done using the {\sc mrqmin}
routine (using the Levenberg-Marquardt method) from Press et al. (1992) 
written in Fortran77, which minimises
\begin{equation} 
  \chi^2= \sum_{i=1}^{i=n} (F_i - F(t_i))^2 / (\sigma_{\rm F_i})^2,
\end{equation} 
with $F_i$ the measurement at time $t_{\rm i}$ which has an error bar $\sigma_{\rm F_i}$.
Also the reduced $\chi^2$ is defined:
\begin{equation} 
  \chi_{\rm r}^2 = \frac{\chi^2}{ (n-p)} 
\end{equation} 
and the quantity
\begin{equation} 
  {\rm BIC} = \chi^2 + (p + 1)\; \ln (n), 
\end{equation} 
where $p = 2 N + 2$ is the number of free parameters ($p = 2 N + 1$
when fitting the RV and $K$ light curve).  As the number $N$ of
harmonics to be fitted to the data is a priori not known, one could
obtain ever better fits (lower $\chi^2$) by increasing $N$.  The
Bayesian information criterion (Schwarz 1978) is a formalism that
penalises this, and $N$ (for the $V$, $K$ and RV curve independently)
is chosen such that BIC reaches a minimum. The number of harmonics
used varies between 3 and 10 in the optical, 1 and 5 in the NIR, and 2
and 9 for the RV curves.

Given the analytical form of Eq.~1, the radial velocity curve can be
integrated exactly to obtain the variation in radius as a function of
time (phase):
\begin{displaymath}
\Delta R (t, \delta \theta)  = -p \; \int_{t_0}^{t + P \delta \theta} (v_{\rm R} - \gamma) \, d t,
\end{displaymath}
where $\gamma$ is the systemic velocity, $v_{\rm R}$ the radial
velocity, $p$ the projection factor and $\delta \theta$ allows for a
phase shift between the RV curve and the angular diameter variations determined via the SB relation.

Then, the equation
\begin{equation} 
  \theta (t) = 9.3038\, {\rm mas} \; \left( \frac{ R_0 + \Delta R (t, \delta \theta) }{ d } \right)
\label{eq-fit}
\end{equation} 
is fitted with $\theta$ the angular diameter in mas, $R_0$ the stars
radius in solar radii and $d$ the distance in pc.  The fitting is done
using the linear bi-sector (using the code SIXLIN from Isobe et al. 1990) 
as used and preferred by e.g. Storm et al. (2004), Barnes et al. (2005) 
and Gieren et al. (2005). In some cases a phase range around 0.85-0.95
was excluded from the fit.


In G07 the $p$-factor was determined chiefly using six Cepheids
with interferometrically measured angular diameter variations and HST
based accurate parallaxes (Benedict et al. 2007). It was
concluded that, formally, there was no need for a period-dependent
$p$-factor, and $p$ = 1.27 $\pm$ 0.05 fitted the available data.  On
the other hand, a moderate dependence of $p \sim - 0.03 \log P$ as
used in the Gieren et al. papers (1993, 1997, 1998, Storm et al. 2004,
Barnes et al. 2003), or $p \sim - 0.064 \log P$ as advocated by
Nardetto et al. (2007) are also consistent with the available data.

The Nardetto et al. (2007) theoretical investigation suggest that
there is a difference between the $p$-factor to be used with wide-band
interferometry (like in G07) and with RV data (when applying the
SB technique as in the present study).  For $\delta$ Cep this
difference is of the order of 0.06 (Nardetto et al. 2004, 2007).

In the present paper, two $p$-factors will be considered: $p$ = 1.33
(the value found in G07 corrected by 0.06), and $p = 1.376 - 0.064 \log P$ 
from Nardetto et al. (2007).  The derived distances depend linearly on
the adopted value of $p$.

An SB relation can be defined as follows (see van Belle 1999):
\begin{equation}
\theta_o = \theta \times 10^{(m_1/5)},
\end{equation}
where $\theta$ is the LD angular diameter (in mas), and
$m_1$ a de-reddened magnitude (for example, $V_0$). The logarithm of
this quantity (the zero magnitude angular diameter) is plotted against a de-reddened colour (for example, $(V-K)_0$), 
\begin{equation}
\log \theta_0 = a \times (m_2 - m_3) + b.
\end{equation}

G07 analysed the available datasets of Cepheids with interferometrically 
determined angular diameters over the light curve and derived a SB relation. 
In a recent paper, Feast et al. (2008) presented new NIR data for zeta Gem and X Sgr. 
Taking this new data into account, Eq.~9 in G07 becomes
\begin{equation}
   \log \theta_0 = (0.2724 \pm 0.0045) \; (V-K)_0 + (0.5283 \pm 0.0091)
\end{equation}
which is the form of the the SB relation used in this paper.


Colour excess values $E(B-V)$ used to de-redden the $V$ and $K$ magnitudes
have been taken from the compilation by Fouqu\'e et al. (2007;
hereafter F07).  Reddening ratios of $R_{\rm V}$= 3.3 and $A_{\rm
  K}/A_{\rm V}$= 0.0909 have been adopted, as in G07.
For the stars already studied in G07 (beta Dor, l Car, zeta Gem, W Sgr,
Y Sgr, del Cep, FF Aql, T Vul, RT Aur), for consistency, the $E(B-V)$
values from G07 have been taken, which are based on the $A_{\rm V}$
values in Benedict et al. (2007) and the reddening ratio of 3.3.  The
resulting $E(B-V)$ values do not differ significantly from the values
listed in F07.

\section{Binary Cepheids}
\label{sect-bin}
 
A number of stars in the sample are known or suspected spectroscopic
binaries. In order to apply the SB technique outlined in the previous
section, the RV data have to be corrected for the binary motion.

The database on Cepheids in binary
systems\footnote{http://www.konkoly.hu/CEP/intro.html} (Szabados 2003)
is most useful as it lists published orbital parameters and references.
For some stars, more RV data have become available than was used in the last
paper that published an orbit and so it was decided to rederive the 5
orbital parameters (in the case of the eccentricity, $\ln e$ was the
parameter actually fitted).

A weighted fit to the RV curve was performed using MRQMIN with 
(2 N + 2) + 5 free parameters (sometimes less when e.g. the
eccentricity was fixed), with $N$ the number of Fourier terms in the
pulsation curve. This procedure does not take into account the
possibility that the period also changed over the time spanned by the
RV data.

As the data sometimes spans more than a hundred years of data with
possible shifts in zero points and varying intrinsic accuracies, the
fitting was done as follows. Initially all data points were given the
same error in RV, and the fit was made.  A mean offset and rms per
dataset between data and model was calculated. If the offset was more
than half the rms, the offset was applied. A second fit was made. The
binary motion was subtracted and the datasets compared as outlined
above in order to assign realistic error bars per dataset. The fit to
the binary + pulsation RV curve was repeated to arrive at the final fit.

The orbital elements are listed in Table~\ref{Tab-Bin}. For most of the known
spectroscopic binaries the derived orbital parameters are close to
literature values, cf. Evans et al. (1990) for FF Aql, Imbert (1996)
for Z Lac and U Vul, Evans et al. (1999) for T Mon, Petterson et al. (2004) 
for S Mus and V350 Sgr, Benedict et al. (2007) for W Sgr,
Gorynya et al. (1995) for S Sge, Evans (1988) for SU Cyg.

The only exceptions are VZ Cyg and SU Cas. For VZ Cyg, Gorynya et
al. (1995) quote an orbital period of 725 days, a velocity amplitude
of 6.5 \ks\ and an eccentricity of 0.05, while Rastorgouev et
al. (1997) give very different values of $P$= 1483d, $K$= 15.1 \ks, $e$= 0.74.
After pre-whitening the RV curve with the pulsation period. a Fourier
analysis using Period04 indicates a period of 2080 days with an
amplitude of 3.3 \ks. Using a larger dataset than the previous papers
the eccentric orbit could not be confirmed, and Table~\ref{Tab-Bin}
lists the derived elements for $e$= 0, which results in an orbital
period of 2092 days.
For SU Cas, Gorynya et al. (1996) quote a period of 408 days, velocity
amplitude of 3 \ks\ and $e$= 0.43, while Rastorgouev et al. (1997)
give $P$= 407d, $K$= 5.2 \ks, $e$= 0.73. After pre-whitening the RV curve
with the pulsation period, a Fourier analysis using Period04 indicates
a period of 408 days with an amplitude of 1.2 \ks. Using the larger
dataset than the previous works, the eccentric orbit could not be
confirmed, and Table~\ref{Tab-Bin} lists the derived elements for $e$= 0, 
which results in an orbital period of 407 days and a small velocity amplitude.

For four stars no orbital elements seem to have been published before.
For V496 Aql, Szabados (1989) suggested periods of 1200, 1780, 2700,
3600, 5370, or 10750 days.  After pre-whitening the RV curve with the
pulsation period a Fourier analysis using Period04 indicates
periods of 529, 1355, 1825, 4065 days with an amplitude of 2.6-3.4 \ks.

For a fixed velocity amplitude of 3 \ks\ and zero eccentricity the
period parameter space was searched for the best fit. This was found
for a period of 1331 days, closely followed by a fit with a period of
1769 days. Other periods resulted in much poorer fits. Although
clearly more RV data are needed to better determine the orbital parameters, 
the RV curve was corrected using the parameters listed in Table~\ref{Tab-Bin}.

For X Sgr a period significantly longer than the 507.25 days suggested
by Szabados (1990) is derived even though only a few additional RV have
become available since. The orbital elements in Table~\ref{Tab-Bin}
assume zero eccentricity.

For S Nor, Szabados (1989) suggested periods of 3300 or 6500 days. 
After pre-whitening the RV curve with the pulsation period a Fourier
analysis using Period04 indicates a period of 7910 days with an
amplitude of 3.2 \ks. Fixing the eccentricity at zero, the best
fitting period is found to be near 3600 days. Although more RV
data are needed to better determine the orbital parameters, the RV
curve was corrected using the parameters listed in Table~\ref{Tab-Bin}.

For XX Cen, Szabados (1990) suggested a period of 909 days which is largely confirmed.
Although clearly more RV data are needed to better determine the
orbital parameters, the RV curve was corrected using the parameters
listed in Table~\ref{Tab-Bin} which have been derived for zero eccentricity.

Finally, for Y Oph the suggestion in Evans \& Lyons (1986) that there
is little evidence for orbital motion is confirmed using significantly
more data than they had at their disposal. 
For LS Pup, Szabados (1996) suggested it to be a binary but no new RV
data have been published for this star in the last ten years and no
correction for binary motion has been attempted.
For Y Sgr, Szabados (1989) suggested a binary period of order
10000-12000 days (and longer could not be excluded).  Assuming zero
excentricity, several oribital periods longer than 10000 days were
tried, but no convincing solution could be found.
For BF Oph, Szabados (1989) suggested a binary period of 4420 $\pm$ 80
days. After pre-whitening the RV curve with the pulsation period, a
Fourier analysis using Period04 indicates a period of 4085 days with a
velocity amplitude near 3 \ks. The least-squares fitting to the RV
data then revealed an alternative good fit with a period of 24700
$\pm$ 2900 days with an amplitude of 2.5 \ks. In the end the RV data
of BF Oph was not corrected.

\begin{table*} 
\caption{Derived orbital parameters of binary Cepheids. Quantities without error bar have been fixed.} 
\begin{tabular}{rcccccc} \hline \hline 
Name     &       $K$  (\ks) &               $e$        &   $\omega$ (degr) &  $T_0$  (JD-2400000)  &       Period (d)    \\  
\hline 
FF Aql   &  4.91 $\pm$ 0.07 & 0.027 $^{+0.041}_{-0.016}$ & 319   $\pm$ 45   &  45437   $\pm$ 178   &  1432.4  $\pm$   1.1  \\
S Mus    & 14.78 $\pm$ 0.10 & 0.086 $^{+0.008}_{-0.007}$ & 194   $\pm$  4   &  48575   $\pm$   5   &   505.15 $\pm$   0.08 \\
S Nor    &  2.53 $\pm$ 0.35 & 0                        & 0                &  45639   $\pm$  67   &  3584    $\pm$  33    \\
S Sge    & 15.62 $\pm$ 0.06 & 0.238 $^{+0.005}_{-0.005}$ & 202.7 $\pm$  1.1 &  48010.7 $\pm$   1.9 &   675.72 $\pm$   0.04 \\
SU Cas   &  0.98 $\pm$ 0.08 & 0                        & 0                &  50279   $\pm$   6   &   406.76 $\pm$   0.04 \\ 
SU Cyg   & 29.83 $\pm$ 0.15 & 0.350 $^{+0.004}_{-0.003}$ & 223.85 $\pm$ 0.85 & 43766.2 $\pm$   1.1 &   549.24 $\pm$   0.02 \\ 
T Mon    &  8.42 $\pm$ 0.19 & 0.414 $^{+0.013}_{-0.013}$ & 204   $\pm$  3   &  49300   $\pm$ 143   & 32449    $\pm$ 726    \\
U Vul    &  3.64 $\pm$ 0.44 & 0.675 $^{+0.033}_{-0.031}$ & 353   $\pm$  4   &  44800   $\pm$  16   &  2510    $\pm$   2.8  \\
V350 Sgr & 11.21 $\pm$ 0.12 & 0.405 $^{+0.015}_{-0.014}$ & 279   $\pm$  2   &  47594   $\pm$   9   &  1482    $\pm$   2.4  \\ 
V496 Aql &  3.0             & 0                        & 0                &  45606   $\pm$  25   &  1331    $\pm$   6.5  \\
VZ Cyg &    3.46 $\pm$ 0.19 & 0                        & 0                &  45002   $\pm$ 50    &  2092    $\pm$  21    \\ 
W Sgr    &  1.43 $\pm$ 0.06 & 0.153 $^{+0.046}_{-0.035}$ & 331   $\pm$ 16   &  48380   $\pm$  79   &  1651    $\pm$   8.8  \\
X Sgr    &  2.30 $\pm$ 0.27 & 0                        & 0                &  48208   $\pm$  19   &   573.6  $\pm$   0.6  \\
XX Cen   &  4.47 $\pm$ 0.28 & 0                        & 0                &  44861   $\pm$   8   &   924.1  $\pm$   1.1  \\
Z Lac    & 10.44 $\pm$ 0.10 & 0.025 $^{+0.012}_{-0.008}$ & 344   $\pm$ 21   &  46582   $\pm$  22   &   382.63 $\pm$   0.10 \\

\hline 
\end{tabular} 
\label{Tab-Bin} 
\end{table*}

\section{Results}

\subsection{Distances and magnitudes}

Table~\ref{Tab-dist} lists the distances, radii and absolute magnitudes obtained.  
The table also lists the adopted $E(B-V)$ and error bar, the derived
period, and the $p$-factor following $p = 1.376 - 0.064 \log P$ from
Nardetto et al. (2007). The distances, radii and errors scale directly
with $p$, so the values for $p$ = 1.33 are not repeated. The error in 
the period is a few units in the last decimal place. For the
derived quantities two error bars are quoted. For the distance and
radius the first error bar listed is the error in the fit, and for the
absolute magnitudes the error due to the error in distance and
$E(B-V)$. The second error quoted is based on a Monte Carlo simulation
where new datasets are generated based on the error bar in each
individual $V$, $K$, RV measurement, and analysed taking an $E(B-V)$
value randomly drawn from a Gaussian distribution based on the listed
mean value and 1$\sigma$ error bar. The second error quoted is the
1$\sigma$ dispersion in the derived quantities.

\begin{figure} 

\begin{minipage}{0.49\textwidth}
\resizebox{\hsize}{!}{\includegraphics{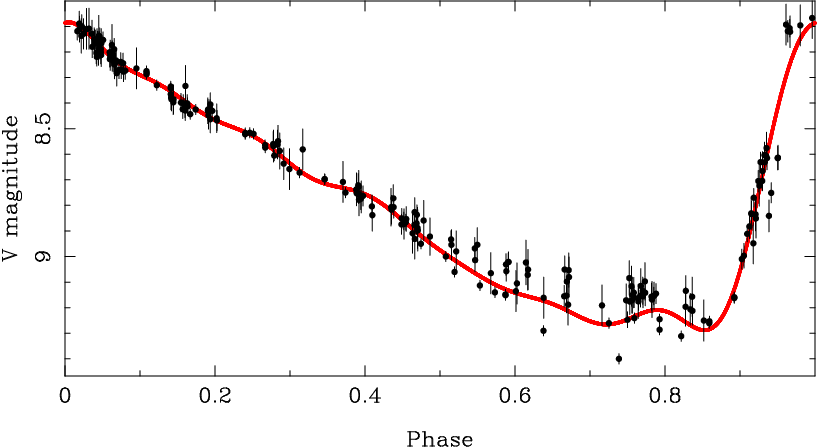}}
\end{minipage}
\begin{minipage}{0.49\textwidth}
\resizebox{\hsize}{!}{\includegraphics{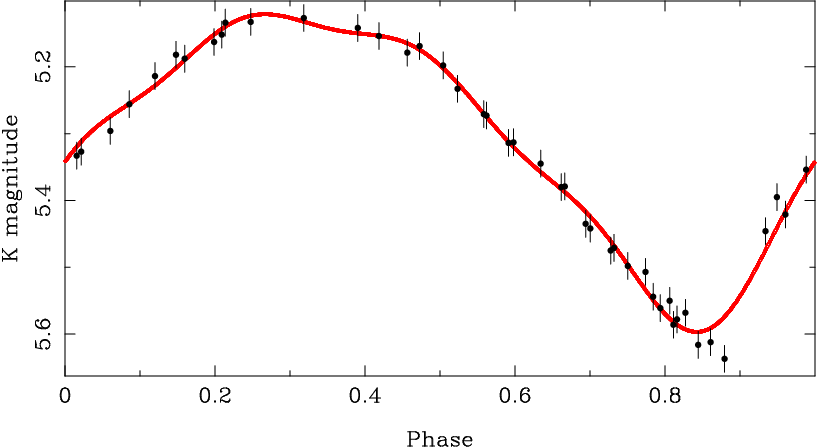}}
\end{minipage}
\begin{minipage}{0.49\textwidth}
\resizebox{\hsize}{!}{\includegraphics{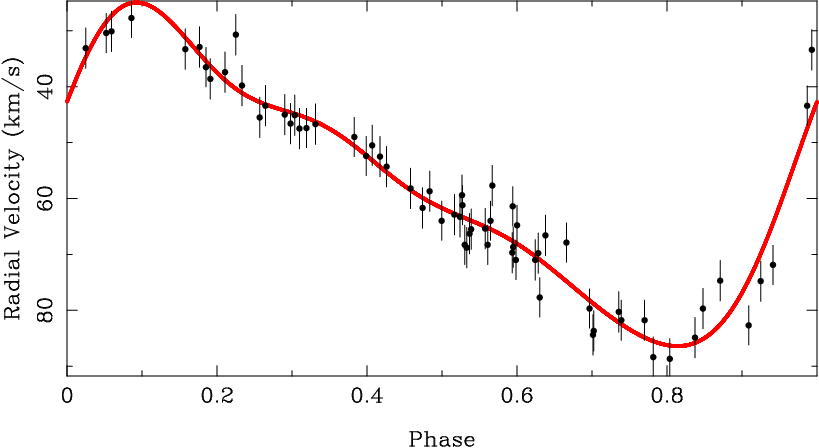}}
\end{minipage}

\caption[]{ 
The phased curves in $V$, $K$ and RV are shown for AQ Pup. Data
points are shown with errors bars and the line shows the harmonic fit. 
The fits to all stars in the sample are available from the author upon request.
} 
\label{Fig-AQPUP1} 
\end{figure}

\begin{figure} 

\begin{minipage}{0.49\textwidth}
\resizebox{\hsize}{!}{\includegraphics{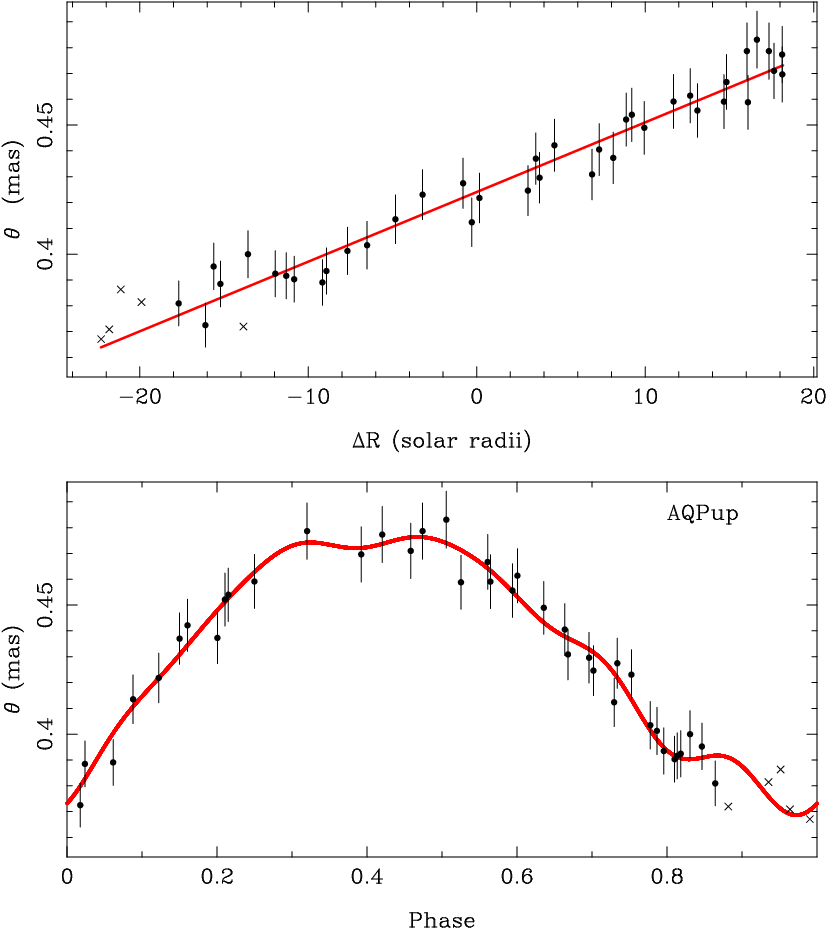}}
\end{minipage}

\caption[]{ 
For AQ Pup, the top panel shows the linear-bisector fit
to the angular diameter as a function of radial displacement.  The
bottom panel shows the angular diameter against phase. Crosses
represent datapoints not considered in the fit. The fits to all stars
in the sample are available from the author upon request.
} 
\label{Fig-AQPUP2} 
\end{figure}

Figure~\ref{Fig-AQPUP1} illustrates the fit to the $V$, $K$, and RV
curve for AQ Pup, while Fig.~\ref{Fig-AQPUP2} shows the variation of
the angular diameter against phase and the change in angular diameter
derived from the SB relation against the change in radius from
integration of the RV curve from which the distance is derived (see
Eq.~\ref{eq-fit}). Figures similar to \ref{Fig-AQPUP1} and
\ref{Fig-AQPUP2} for all stars in the sample are available from the author.

For ten stars the best available distance comes from the HST-based
parallax (Benedict et al. 2007). Although a BW-type analysis has been
carried out for these stars (also see G07), only the mean $V$ and
$K$ magnitudes derived here have been used together with the parallax
from Benedict et al. to derive the absolute magnitudes listed in
Table~\ref{Tab-dist1}.

F07 studied the calibration of the Galactic $PL$-relation in
various photometric bands using a set of 59 calibrating stars, with
distances from HST and revised Hipparcos parallaxes, interferometric
Baade-Wesselink and infrared SB parallaxes and Main-Sequence fitting.

Of the  58  stars in Table~\ref{Tab-dist}, 57 have a distance quoted in F07. 
The ratio of the distance quoted in Table~\ref{Tab-dist} and that in
Table~6 of finally adopted parallaxes in F07 is 1.02 $\pm$ 0.11 (minus
one outlier, X Lac). The absolute difference in distance between the
two estimates is 1.1 times the combined error bar. Regarding the
absolute magnitudes quoted in their Table~7, one difference w.r.t. the
values in Table~\ref{Tab-dist} is due to the fact that F07 use
published intensity-mean values, while in the present paper these have
been derived in a self-consistent way from the Fourier fit to the data.








\begin{table*} 
\setlength{\tabcolsep}{1.3mm}

\caption{Distances, radii and absolute magnitudes from the BW-analysis} 


\begin{tabular}{rcrcrrcc} \hline \hline 

Name       &  $E(B-V)$ & Period (d) & $p$ &  $d$       &   $R$ (\rsol)    & $M_{\rm V}$ & $M_{\rm K}$  \\  
\hline

AQ Pup & 0.518 $\pm$ 0.010 & 30.0980   & 1.281 & 3292.2 $\pm$ 80.0 $\pm$ 125.8 & 149.7  $\pm$ 3.7 $\pm$ 6.0 & -5.51 $\pm$ 0.06 $\pm$ 0.09 & -7.44 $\pm$ 0.05 $\pm$ 0.09 \\
BB Sgr & 0.281 $\pm$ 0.009 &  6.637111 & 1.323 & 768.7 $\pm$ 25.0 $\pm$  25.7 & 46.9  $\pm$ 1.5 $\pm$ 1.5 & -3.40 $\pm$ 0.08 $\pm$ 0.08 & -5.00 $\pm$ 0.07 $\pm$ 0.07 \\
BF Oph & 0.235 $\pm$ 0.010 &  4.067677 & 1.337 & 727.7 $\pm$ 32.6 $\pm$  28.9 & 31.9  $\pm$ 1.4 $\pm$ 1.3 & -2.72 $\pm$ 0.10 $\pm$ 0.09 & -4.20 $\pm$ 0.10 $\pm$ 0.09 \\
BG Lac & 0.300 $\pm$ 0.016 &  5.33192  & 1.329 & 1700.9 $\pm$ 24.3 $\pm$  38.5 & 39.9  $\pm$ 0.6 $\pm$ 0.9 & -3.24 $\pm$ 0.06 $\pm$ 0.08 & -4.70 $\pm$ 0.03 $\pm$ 0.05 \\
BN Pup & 0.416 $\pm$ 0.018 & 13.67243  & 1.303 & 3773.4 $\pm$ 94.6 $\pm$ 145.9 & 80.2  $\pm$ 2.0 $\pm$ 3.2 & -4.30 $\pm$ 0.08 $\pm$ 0.11 & -6.08 $\pm$ 0.05 $\pm$ 0.09 \\
CD Cyg & 0.493 $\pm$ 0.015 & 17.07406  & 1.297 & 2336.7 $\pm$ 64.1 $\pm$  60.5 & 88.3  $\pm$ 2.4 $\pm$ 2.2 & -4.46 $\pm$ 0.08 $\pm$ 0.08 & -6.31 $\pm$ 0.06 $\pm$ 0.06 \\
CV Mon & 0.722 $\pm$ 0.022 &  5.378662 & 1.329 & 1475.1 $\pm$ 60.7 $\pm$  53.9 & 36.7  $\pm$ 1.5 $\pm$ 1.3 & -2.92 $\pm$ 0.11 $\pm$ 0.12 & -4.48 $\pm$ 0.09 $\pm$ 0.08 \\
DT Cyg & 0.042 $\pm$ 0.011 &  2.499189 & 1.351 & 603.3 $\pm$ 75.7 $\pm$  66.5 & 34.9  $\pm$ 4.4 $\pm$ 3.9 & -3.26 $\pm$ 0.26 $\pm$ 0.29 & -4.48 $\pm$ 0.26 $\pm$ 0.28 \\
eta Aql & 0.130 $\pm$ 0.009 & 7.176813 & 1.321 & 261.0 $\pm$ 6.7 $\pm$   7.1 & 49.3  $\pm$ 1.3 $\pm$ 1.3 & -3.59 $\pm$ 0.06 $\pm$ 0.07 & -5.14 $\pm$ 0.05 $\pm$ 0.06 \\
FM Aql & 0.589 $\pm$ 0.012 &  6.114274 & 1.326 & 1141.2 $\pm$ 38.8 $\pm$  37.4 & 55.1  $\pm$ 1.9 $\pm$ 1.8 & -3.93 $\pm$ 0.08 $\pm$ 0.09 & -5.40 $\pm$ 0.07 $\pm$ 0.07 \\
FN Aql & 0.483 $\pm$ 0.010 &  9.48163 & 1.313 & 1217.5 $\pm$ 28.8 $\pm$  25.9 & 52.2  $\pm$ 1.2 $\pm$ 1.1 & -3.62 $\pm$ 0.06 $\pm$ 0.06 & -5.21 $\pm$ 0.05 $\pm$ 0.04 \\
KN Cen & 0.797 $\pm$ 0.091 & 34.0498 & 1.278 & 3973.5 $\pm$ 93.4 $\pm$ 161.9 & 177.1  $\pm$ 4.2 $\pm$ 4.5 & -5.71 $\pm$ 0.30 $\pm$ 0.35 & -7.74 $\pm$ 0.06 $\pm$ 0.11 \\
KQ Sco & 0.869 $\pm$ 0.021 & 28.6958 & 1.283 & 2655.2 $\pm$ 56.4 $\pm$ 105.8 & 160.9  $\pm$ 3.4 $\pm$ 6.2 & -5.13 $\pm$ 0.08 $\pm$ 0.12 & -7.43 $\pm$ 0.05 $\pm$ 0.09 \\
LS Pup & 0.461 $\pm$ 0.015 & 14.14725 & 1.302 & 4919.8 $\pm$ 91.3 $\pm$ 157.8 & 85.4  $\pm$ 1.6 $\pm$ 2.7 & -4.49 $\pm$ 0.06 $\pm$ 0.09 & -6.24 $\pm$ 0.04 $\pm$ 0.07 \\
RS Pup & 0.457 $\pm$ 0.009 & 41.457 & 1.272 & 1765.4 $\pm$ 33.9 $\pm$  65.2 & 174.2  $\pm$ 3.4 $\pm$ 6.8 & -5.67 $\pm$ 0.05 $\pm$ 0.09 & -7.71 $\pm$ 0.04 $\pm$ 0.08 \\
RU Sct & 0.921 $\pm$ 0.012 & 19.7032 & 1.293 & 2043.5 $\pm$ 78.3 $\pm$  60.9 & 107.6  $\pm$ 4.1 $\pm$ 3.2 & -5.04 $\pm$ 0.09 $\pm$ 0.08 & -6.76 $\pm$ 0.08 $\pm$ 0.07 \\
RY Sco & 0.718 $\pm$ 0.018 & 20.32134 & 1.292 & 1185.1 $\pm$ 26.0 $\pm$  56.8 & 94.1  $\pm$ 2.1 $\pm$ 4.4 & -4.68 $\pm$ 0.08 $\pm$ 0.13 & -6.44 $\pm$ 0.05 $\pm$ 0.11 \\
RY Vel & 0.547 $\pm$ 0.010 & 28.1240 & 1.283 & 2178.5 $\pm$ 28.1 $\pm$  95.3 & 118.6  $\pm$ 1.5 $\pm$ 5.0 & -5.08 $\pm$ 0.04 $\pm$ 0.10 & -6.92 $\pm$ 0.03 $\pm$ 0.09 \\
RZ Vel & 0.299 $\pm$ 0.010 & 20.3996 & 1.292 & 1472.3 $\pm$ 17.9 $\pm$  27.1 & 103.8  $\pm$ 1.3 $\pm$ 1.8 & -4.67 $\pm$ 0.04 $\pm$ 0.06 & -6.61 $\pm$ 0.03 $\pm$ 0.04 \\
S Mus & 0.212 $\pm$ 0.016  &  9.65997 & 1.313 & 787.7 $\pm$ 29.5 $\pm$  24.2 & 59.0  $\pm$ 2.2 $\pm$ 1.8 & -4.03 $\pm$ 0.10 $\pm$ 0.09 & -5.53 $\pm$ 0.08 $\pm$ 0.07 \\
S Nor & 0.179 $\pm$ 0.009  &  9.75425 & 1.313 & 819.8 $\pm$ 21.9 $\pm$  23.7 & 60.0  $\pm$ 1.6 $\pm$ 1.8 & -3.71 $\pm$ 0.06 $\pm$ 0.07 & -5.46 $\pm$ 0.06 $\pm$ 0.06 \\
S Sge & 0.100 $\pm$ 0.010  &  8.38207 & 1.317 & 662.9 $\pm$ 17.1 $\pm$  14.0 & 55.5  $\pm$ 1.4 $\pm$ 1.2 & -3.80 $\pm$ 0.06 $\pm$ 0.06 & -5.37 $\pm$ 0.06 $\pm$ 0.05 \\
SS Sct & 0.325 $\pm$ 0.010 &  3.671331 & 1.340 & 959.1 $\pm$ 90.6 $\pm$  71.7 & 30.7  $\pm$ 2.9 $\pm$ 2.4 & -2.76 $\pm$ 0.20 $\pm$ 0.18 & -4.15 $\pm$ 0.20 $\pm$ 0.17 \\
SU Cas & 0.259 $\pm$ 0.010 &  1.949329 & 1.357 & 356.8 $\pm$ 18.8 $\pm$  22.9 & 23.9  $\pm$ 1.3 $\pm$ 1.6 & -2.65 $\pm$ 0.12 $\pm$ 0.16 & -3.71 $\pm$ 0.11 $\pm$ 0.15 \\
SU Cyg & 0.098 $\pm$ 0.014 &  3.845552 & 1.339 & 842.1 $\pm$ 62.0 $\pm$  39.0 & 32.9  $\pm$ 2.4 $\pm$ 1.5 & -3.05 $\pm$ 0.16 $\pm$ 0.11 & -4.34 $\pm$ 0.15 $\pm$ 0.10 \\
S Vul  & 0.727 $\pm$ 0.042 & 68.714 & 1.258 & 4430.4 $\pm$ 147.7 $\pm$ 205.7 & 304.8  $\pm$ 10.2 $\pm$ 12.7 & -6.65 $\pm$ 0.16 $\pm$ 0.19 & -8.85 $\pm$ 0.07 $\pm$ 0.10 \\
SV Vul & 0.461 $\pm$ 0.022 & 45.028 & 1.270 & 2449.9 $\pm$ 57.8 $\pm$  58.2 & 211.1  $\pm$ 5.0 $\pm$ 7.8 & -6.21 $\pm$ 0.09 $\pm$ 0.10 & -8.15 $\pm$ 0.05 $\pm$ 0.06 \\
SW Vel & 0.344 $\pm$ 0.010 & 23.4395 & 1.288 & 2285.6 $\pm$ 25.5 $\pm$  54.9 & 104.1  $\pm$ 1.2 $\pm$ 2.5 & -4.73 $\pm$ 0.04 $\pm$ 0.07 & -6.65 $\pm$ 0.02 $\pm$ 0.05 \\
SX Vel & 0.236 $\pm$ 0.012 &  9.55042 & 1.313 & 1788.9 $\pm$ 70.4 $\pm$ 121.5 & 54.5  $\pm$ 2.1 $\pm$ 3.7 & -3.74 $\pm$ 0.09 $\pm$ 0.15 & -5.33 $\pm$ 0.08 $\pm$ 0.15 \\
SZ Aql & 0.537 $\pm$ 0.017 & 17.13977 & 1.297 & 1960.2 $\pm$ 24.3 $\pm$  47.3 & 96.4  $\pm$ 1.2 $\pm$ 2.3 & -4.54 $\pm$ 0.06 $\pm$ 0.09 & -6.46 $\pm$ 0.03 $\pm$ 0.06 \\
SZ Tau & 0.295 $\pm$ 0.011 &  3.148921 & 1.344 & 525.3 $\pm$ 12.1 $\pm$  21.6 & 33.9  $\pm$ 0.8 $\pm$ 1.4 & -3.05 $\pm$ 0.06 $\pm$ 0.10 & -4.38 $\pm$ 0.05 $\pm$ 0.09 \\
T Mon & 0.181 $\pm$ 0.011  & 27.0318 & 1.284 & 1239.0 $\pm$ 20.0 $\pm$  19.6 & 125.7  $\pm$ 2.0 $\pm$ 1.8 & -4.88 $\pm$ 0.05 $\pm$ 0.06 & -6.98 $\pm$ 0.03 $\pm$ 0.04 \\
TT Aql & 0.438 $\pm$ 0.011 & 13.75502 & 1.303 & 1025.1 $\pm$ 22.1 $\pm$  27.2 & 82.0  $\pm$ 1.8 $\pm$ 2.1 & -4.32 $\pm$ 0.06 $\pm$ 0.07 & -6.12 $\pm$ 0.05 $\pm$ 0.06 \\
T Vel & 0.289 $\pm$ 0.010  &  4.639811 & 1.333 & 942.2 $\pm$ 9.2 $\pm$  27.5 & 33.9  $\pm$ 0.3 $\pm$ 1.0 & -2.78 $\pm$ 0.04 $\pm$ 0.07 & -4.31 $\pm$ 0.02 $\pm$ 0.06 \\
U Car & 0.265 $\pm$ 0.010  & 38.8229 & 1.274 & 1684.2 $\pm$ 51.9 $\pm$  42.8 & 170.0  $\pm$ 5.3 $\pm$ 4.5 & -5.66 $\pm$ 0.07 $\pm$ 0.07 & -7.68 $\pm$ 0.07 $\pm$ 0.06 \\
U Nor & 0.862 $\pm$ 0.024  & 12.64418 & 1.305 & 1438.3 $\pm$ 26.0 $\pm$  32.0 & 78.3  $\pm$ 1.4 $\pm$ 1.6 & -4.37 $\pm$ 0.09 $\pm$ 0.10 & -6.05 $\pm$ 0.04 $\pm$ 0.05 \\
U Sgr & 0.403 $\pm$ 0.009  &  6.745306 & 1.323 & 541.0 $\pm$ 23.9 $\pm$  22.4 & 42.9  $\pm$ 1.9 $\pm$ 1.8 & -3.28 $\pm$ 0.10 $\pm$ 0.10 & -4.83 $\pm$ 0.09 $\pm$ 0.10 \\
UU Mus & 0.399 $\pm$ 0.015 & 11.63613 & 1.308 & 2906.1 $\pm$ 63.2 $\pm$ 112.0 & 64.6  $\pm$ 1.4 $\pm$ 2.5 & -3.79 $\pm$ 0.07 $\pm$ 0.10 & -5.60 $\pm$ 0.05 $\pm$ 0.09 \\
U Vul & 0.603 $\pm$ 0.011  &  7.99074 & 1.318 & 638.4 $\pm$ 23.6 $\pm$  21.3 & 50.6  $\pm$ 1.9 $\pm$ 1.7 & -3.86 $\pm$ 0.09 $\pm$ 0.08 & -5.25 $\pm$ 0.08 $\pm$ 0.07 \\
V340 Nor & 0.321 $\pm$ 0.018 & 11.2884 & 1.309 & 1782.4 $\pm$ 101.6 $\pm$ 131.9 & 70.5  $\pm$ 4.0 $\pm$ 5.2 & -3.90 $\pm$ 0.13 $\pm$ 0.18 & -5.76 $\pm$ 0.12 $\pm$ 0.17 \\
V350 Sgr & 0.299 $\pm$ 0.010 &  5.154248 & 1.330 & 807.8 $\pm$ 18.9 $\pm$  38.0 & 35.4  $\pm$ 0.8 $\pm$ 1.6 & -3.02 $\pm$ 0.06 $\pm$ 0.11 & -4.46 $\pm$ 0.05 $\pm$ 0.10 \\
V496 Aql & 0.397 $\pm$ 0.010 &  6.80702 & 1.323 & 937.9 $\pm$ 80.0 $\pm$  81.7 & 45.5  $\pm$ 3.9 $\pm$ 4.0 & -3.40 $\pm$ 0.18 $\pm$ 0.21 & -4.94 $\pm$ 0.18 $\pm$ 0.20 \\
V Car & 0.169 $\pm$ 0.011  &  6.696707 & 1.323 & 908.5 $\pm$ 28.9 $\pm$  47.0 & 38.2  $\pm$ 1.2 $\pm$ 2.0 & -2.96 $\pm$ 0.08 $\pm$ 0.12 & -4.56 $\pm$ 0.07 $\pm$ 0.12 \\
V Cen & 0.292 $\pm$ 0.012  &  5.494010 & 1.329 & 672.6 $\pm$ 16.5 $\pm$  20.7 & 40.2  $\pm$ 1.0 $\pm$ 1.2 & -3.25 $\pm$ 0.07 $\pm$ 0.08 & -4.72 $\pm$ 0.05 $\pm$ 0.07 \\
VW Cen & 0.428 $\pm$ 0.024 & 15.03733 & 1.301 & 3701.4 $\pm$ 66.8 $\pm$ 120.1 & 87.7  $\pm$ 1.6 $\pm$ 2.7 & -3.96 $\pm$ 0.09 $\pm$ 0.12 & -6.15 $\pm$ 0.04 $\pm$ 0.07 \\
VY Car & 0.237 $\pm$ 0.009 & 18.9050 & 1.294 & 1455.1 $\pm$ 34.2 $\pm$  36.6 & 81.6  $\pm$ 1.9 $\pm$ 2.0 & -4.11 $\pm$ 0.06 $\pm$ 0.06 & -6.07 $\pm$ 0.05 $\pm$ 0.06 \\
VZ Cyg & 0.266 $\pm$ 0.011 &  4.86440 & 1.332 & 1773.2 $\pm$ 51.2 $\pm$  54.5 & 37.4  $\pm$ 1.1 $\pm$ 1.2 & -3.13 $\pm$ 0.07 $\pm$ 0.08 & -4.57 $\pm$ 0.06 $\pm$ 0.07 \\
VZ Pup & 0.459 $\pm$ 0.011 & 23.17494 & 1.289 & 4315.0 $\pm$ 86.7 $\pm$ 162.2 & 99.7  $\pm$ 2.0 $\pm$ 4.0 & -4.98 $\pm$ 0.06 $\pm$ 0.10 & -6.63 $\pm$ 0.04 $\pm$ 0.09 \\
WZ Car & 0.370 $\pm$ 0.011 & 23.01521 & 1.289 & 3600.9 $\pm$ 50.8 $\pm$  55.7 & 101.4  $\pm$ 1.4 $\pm$ 1.8 & -4.66 $\pm$ 0.05 $\pm$ 0.06 & -6.59 $\pm$ 0.03 $\pm$ 0.04 \\
WZ Sgr & 0.431 $\pm$ 0.011 & 21.8501 & 1.290 & 1754.1 $\pm$ 29.8 $\pm$  26.8 & 117.6  $\pm$ 2.0 $\pm$ 1.7 & -4.55 $\pm$ 0.05 $\pm$ 0.06 & -6.77 $\pm$ 0.04 $\pm$ 0.04 \\
X Cyg & 0.228 $\pm$ 0.012  & 16.3856 & 1.298 & 1081.0 $\pm$ 25.3 $\pm$  22.7 & 92.8  $\pm$ 2.2 $\pm$ 1.9 & -4.49 $\pm$ 0.06 $\pm$ 0.07 & -6.39 $\pm$ 0.05 $\pm$ 0.05 \\
X Lac & 0.336 $\pm$ 0.011  &  5.44453 & 1.329 & 1507.3 $\pm$ 83.0 $\pm$  70.9 & 44.3  $\pm$ 2.4 $\pm$ 2.1 & -3.59 $\pm$ 0.12 $\pm$ 0.12 & -4.95 $\pm$ 0.12 $\pm$ 0.11 \\
X Pup & 0.402 $\pm$ 0.009  & 25.9657 & 1.285 & 2779.4 $\pm$ 48.0 $\pm$ 133.1 & 117.0  $\pm$ 2.0 $\pm$ 5.8 & -4.95 $\pm$ 0.05 $\pm$ 0.11 & -6.90 $\pm$ 0.04 $\pm$ 0.10 \\
XX Cen & 0.266 $\pm$ 0.011 & 10.95351 & 1.309 & 1412.5 $\pm$ 32.9 $\pm$  31.9 & 58.0  $\pm$ 1.4 $\pm$ 1.2 & -3.77 $\pm$ 0.06 $\pm$ 0.06 & -5.42 $\pm$ 0.05 $\pm$ 0.05 \\
Y Lac & 0.207 $\pm$ 0.016  &  4.323760 & 1.335 & 2523.9 $\pm$ 74.1 $\pm$  86.0 & 41.8  $\pm$ 1.2 $\pm$ 1.4 & -3.52 $\pm$ 0.08 $\pm$ 0.10 & -4.85 $\pm$ 0.06 $\pm$ 0.07 \\
Y Oph & 0.645 $\pm$ 0.015  & 17.12614 & 1.297 & 594.4 $\pm$ 17.5 $\pm$  21.0 & 85.8  $\pm$ 2.5 $\pm$ 3.0 & -4.84 $\pm$ 0.08 $\pm$ 0.10 & -6.33 $\pm$ 0.06 $\pm$ 0.08 \\
YZ Sgr & 0.281 $\pm$ 0.010 &  9.55376 & 1.313 & 1102.4 $\pm$ 53.9 $\pm$  90.8 & 57.2  $\pm$ 2.8 $\pm$ 4.8 & -3.77 $\pm$ 0.11 $\pm$ 0.19 & -5.41 $\pm$ 0.10 $\pm$ 0.19 \\
Z Lac & 0.370 $\pm$ 0.011  & 10.88569 & 1.310 & 1877.5 $\pm$ 37.4 $\pm$  40.1 & 68.4  $\pm$ 1.4 $\pm$ 1.5 & -4.14 $\pm$ 0.06 $\pm$ 0.07 & -5.78 $\pm$ 0.04 $\pm$ 0.05 \\

\hline 

\end{tabular} 
\label{Tab-dist}
\end{table*}

\begin{table*} 

\caption{Absolute magnitudes for the Cepheids with an HST based parallax} 
\begin{tabular}{rccccc} \hline \hline 
Name       &    $E(B-V)$     &      $\pi$      &     $M_{\rm V}$    &    $M_{\rm K}$  \\  
\hline 

beta Dor & 0.076 $\pm$ 0.015 & 3.14 $\pm$ 0.16 & -4.01 $\pm$ 0.12 & -5.58 $\pm$ 0.11  \\
del Cep  & 0.070 $\pm$ 0.010 & 3.66 $\pm$ 0.15 & -3.43 $\pm$ 0.09 & -4.87 $\pm$ 0.09  \\
FF Aql   & 0.194 $\pm$ 0.018 & 2.18 $\pm$ 0.18 & -3.02 $\pm$ 0.15 & -4.32 $\pm$ 0.14  \\
l Car    & 0.158 $\pm$ 0.018 & 2.01 $\pm$ 0.20 & -5.25 $\pm$ 0.22 & -7.43 $\pm$ 0.21  \\
RT Aur   & 0.061 $\pm$ 0.024 & 2.40 $\pm$ 0.19 & -2.82 $\pm$ 0.18 & -4.19 $\pm$ 0.17  \\
T Vul    & 0.103 $\pm$ 0.018 & 1.90 $\pm$ 0.23 & -3.18 $\pm$ 0.26 & -4.43 $\pm$ 0.25  \\
W Sgr    & 0.112 $\pm$ 0.009 & 2.28 $\pm$ 0.20 & -3.90 $\pm$ 0.19 & -5.43 $\pm$ 0.18  \\
X Sgr    & 0.176 $\pm$ 0.030 & 3.00 $\pm$ 0.18 & -3.62 $\pm$ 0.16 & -5.14 $\pm$ 0.13  \\
Y Sgr    & 0.203 $\pm$ 0.012 & 2.13 $\pm$ 0.29 & -3.26 $\pm$ 0.28 & -4.80 $\pm$ 0.28  \\
zeta Gem & 0.018 $\pm$ 0.010 & 2.78 $\pm$ 0.18 & -3.94 $\pm$ 0.14 & -5.65 $\pm$ 0.14  \\

\hline 

\end{tabular} 
\label{Tab-dist1}
\end{table*}

\subsection{$PL(Z)$ relations}

\begin{figure} 

\begin{minipage}{0.49\textwidth}
\resizebox{\hsize}{!}{\includegraphics{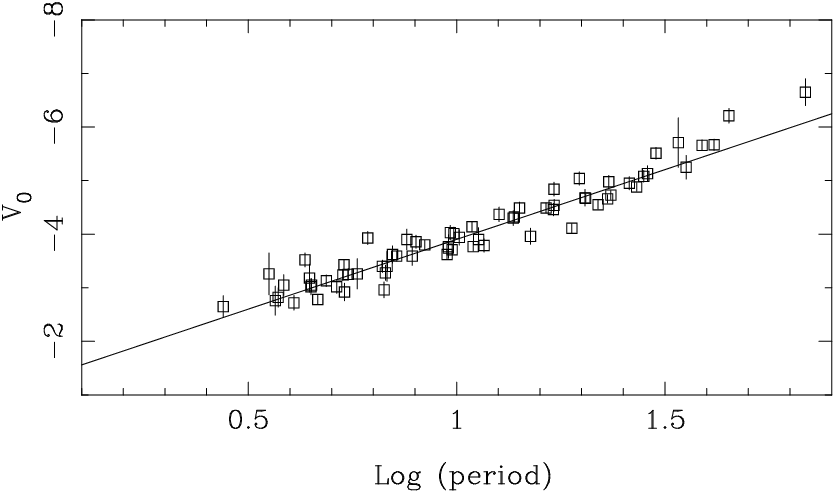}}
\end{minipage}
\begin{minipage}{0.49\textwidth}
\resizebox{\hsize}{!}{\includegraphics{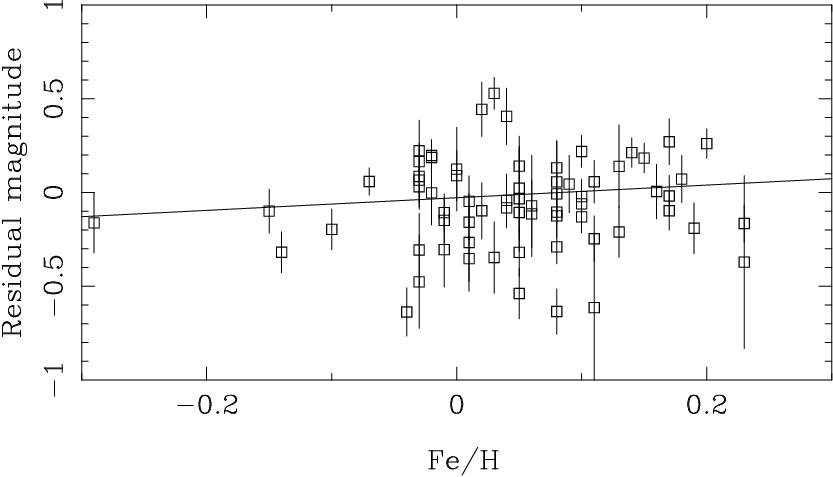}}
\end{minipage}

\caption[]{ 
Period-Luminosity relation in the $V$-band. 
The bottom panel shows the residual plotted
versus metallicity. Lines indicate the fits to the data.
} 
\label{Fig-PLV} 
\end{figure}

\begin{figure} 

\begin{minipage}{0.49\textwidth}
\resizebox{\hsize}{!}{\includegraphics{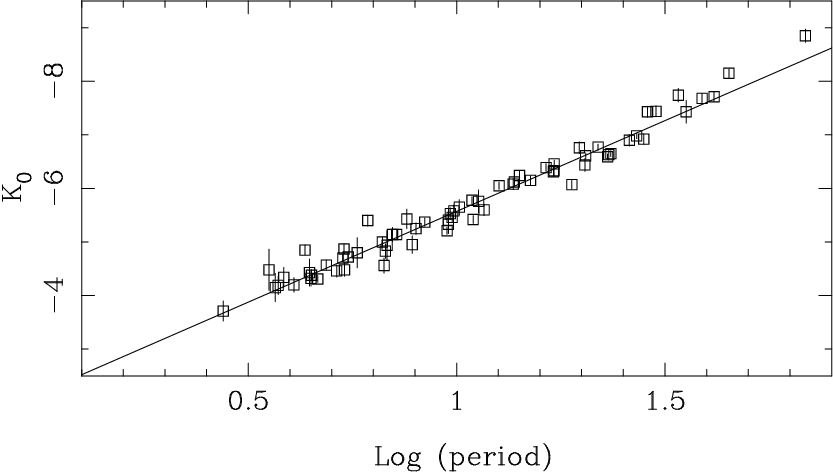}}
\end{minipage}
\begin{minipage}{0.49\textwidth}
\resizebox{\hsize}{!}{\includegraphics{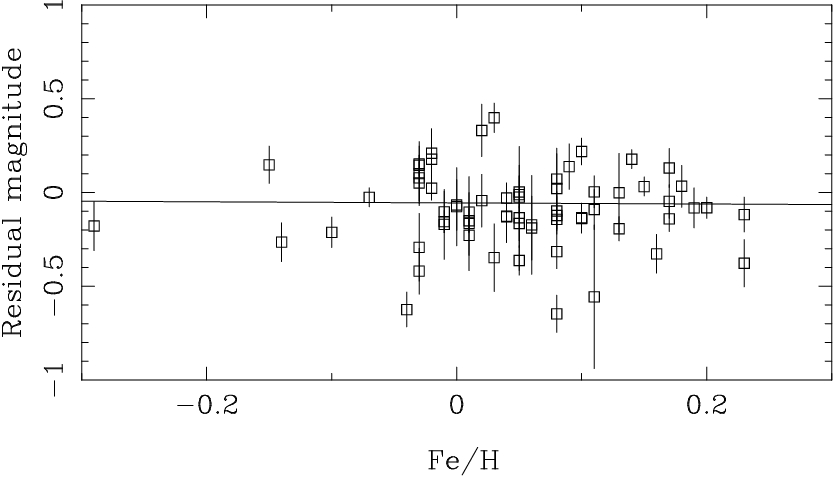}}
\end{minipage}

\caption[]{ 
Period-Luminosity relation in the $K$-band. 
The bottom panel shows the residual plotted
versus metallicity. Lines indicate the fits to the data.
} 
\label{Fig-PLK} 
\end{figure}

Figures~\ref{Fig-PLV} and \ref{Fig-PLK} show the $PL$-relation in
the $V$-band and $K$-band assuming the Nardetto et al. calibration of
the $p$-factor, and in the bottom panel the residual plotted against
metallicity. The first and second error bar quoted for the magnitudes
in Table~\ref{Tab-dist} have been added in quadrature.  The period is
the fundamental one, and the classification into fundamental and first
overtone pulsator is taken from F07. Four stars in the sample are
overtone pulsators (DT Cyg, SU Cas, SZ Tau, X Lac) and their period is
fundamentalised using an iterative procedure using the formula (Alcock
et al. 1995) 
\begin{equation}
P_1/P_0 = 0.720 - 0.027 \log P_0
\end{equation}

The metallicity dependence has been determined in two ways, one by
first fitting a linear $PL$-relation, and fitting the residual with a
linear relation against [Fe/H] (as shown in Figs.~\ref{Fig-PLV} and
\ref{Fig-PLK}), and secondly by making a linear-fit in the two
variables $\log P$ and [Fe/H], as quoted below. The results are
essentially the same.

For the adopted $p$-factor dependence on period the $PLZ$-relation in $V$ becomes

\begin{displaymath}
 M_{\rm V}=  (-2.60 \pm 0.09) \log P + (-1.30 \pm 0.10) 
\end{displaymath}
\begin{equation}
 \hspace{45mm}+ (+0.27 \pm 0.30) {\rm [Fe/H]},
\end{equation}
and in $K$
\begin{displaymath}
 M_{\rm K}=  (-3.38 \pm 0.08) \log P + (-2.19 \pm 0.09) 
\end{displaymath}
\begin{equation}
 \hspace{45mm}+ (-0.11 \pm 0.24) {\rm [Fe/H]}.
\end{equation}
For a constant $p$-factor of 1.33 this would have become
\begin{displaymath}
 M_{\rm V}=  (-2.72 \pm 0.09) \log P + (-1.23 \pm 0.10) 
\end{displaymath}
\begin{equation}
 \hspace{45mm}+ (+0.24 \pm 0.30) {\rm [Fe/H]},
\end{equation}
and in $K$
\begin{displaymath}
 M_{\rm K}=  (-3.50 \pm 0.08) \log P + (-2.09 \pm 0.09) 
\end{displaymath}
\begin{equation}
 \hspace{45mm}+ (-0.10 \pm 0.24) {\rm [Fe/H]}.
\end{equation}
Ignoring the metallicity dependence results in $PL$-relations that
have slopes and zeropoints that differ in the third decimal from the ones quoted above.

Ignoring RS Pup, SV Vul and S Vul which have the longest periods,
  and hence limiting the period to $<$40 days results in

\begin{displaymath}
 M_{\rm V}=  (-2.50 \pm 0.10) \log P + (-1.40 \pm 0.11) 
\end{displaymath}
\begin{equation}
 \hspace{45mm}+ (+0.14 \pm 0.31) {\rm [Fe/H]},
\end{equation}
and in $K$
\begin{displaymath}
 M_{\rm K}=  (-3.29 \pm 0.08) \log P + (-2.28 \pm 0.10) 
\end{displaymath}
\begin{equation}
 \hspace{45mm}+ (-0.01 \pm 0.26) {\rm [Fe/H]}.
\end{equation}
For a constant $p$-factor of 1.33 this would have become
\begin{displaymath}
 M_{\rm V}=  (-2.60 \pm 0.10) \log P + (-1.33 \pm 0.11) 
\end{displaymath}
\begin{equation}
 \hspace{45mm}+ (+0.11 \pm 0.31) {\rm [Fe/H]},
\end{equation}
and in $K$
\begin{displaymath}
f M_{\rm K}=  (-3.41 \pm 0.08) \log P + (-2.18 \pm 0.10) 
\end{displaymath}
\begin{equation}
 \hspace{45mm}+ (-0.02 \pm 0.25) {\rm [Fe/H]}.
\end{equation}

The effect of the negative $\log P$ dependence on $p$ compared to a
constant value is to flatten the $PL$-relation.  
%
The size of the effect is expected to be about 5 $\log$ (1 + 0.064/1.33) 
$\approx$ 0.10 and this is indeed reflected in the fitted slope.  
This change in slope is about equal to the 1$\sigma$ error bar
derived in the slope.

F07 quotes slopes of $-2.77$ $\pm$ 0.08 in $V$ and $-3.37$ $\pm$ 0.06
in $K$ based on 58 stars.  For the LMC they combine data for the OGLE
Cepheid sample (see below) with data from Persson et al. (2004).  They
find a slope of $-2.73 \pm$ 0.03 in $V$, and $-3.23 \pm$ 0.03 in $K$.
The conclusion by F07 is that Galactic and LMC slopes are in good agreement.

There are some issues however that may make a direct comparison
between the LMC and the Galaxy less obvious.  The distribution of the Galactic
Cepheids that define the calibration is more or less uniform in $\log P$ 
(both in F07 and in the present study), while the majority of OGLE
fundamental mode Cepheids has a $\log P$ between 0.4 and 0.7 and a
tail to longer periods (Udalski et al. 1999).  In addition it has been
demonstrated that the $PL$-relation in the LMC is non-linear in $V$
(e.g. Ngeow et al. 2008 and references therein).
Both effects make that the slope based on a least-square fit to the
sample of calibrating Galactic Cepheids cannot be compared directly
to the slope derived for the LMC sample.

To investigate this further we made use of the data kindly made
available by Dr. Soszy\'nski which consists of an OGLE sample of 701
fundamental mode Cepheids whereby the optical light curves have been
used to transform the single-epoch 2MASS data into intensity-mean
values following the prescription in Soszy\'nski et al. (2005). This
is a slightly improved version of the dataset that Dr.  Soszy\'nski
made available to F07. The mean $V$ magnitude and $E(B-V)$ values are
taken from the original OGLE data (Udalski et al. 1999).
The 2MASS data were transformed to the SAAO system following Eq.~10 in Koen et al. (2007).
Removing outliers with iterative 3$\sigma$ clipping the slope in the
LMC $V$-band $PL$-relation becomes $-2.727 \pm 0.033$ based on 664 stars.

In a Monte Carlo simulation a random value of $\log P$ was drawn
between 0.45 (the lower limit of the Galactic Cepheids) and 1.6.  
From the OGLE sample the object closest in $\log P$ was selected and
its $V$ and $K$ magnitude taken.  In this way samples of 66 stars were
generated, and a $PL$-relation fitted. This was repeated a large
number of times. The mean slope and the dispersion in the slope is $-2.69 \pm 0.06$. 

In the $K$-band, after removing outliers, the $PL$-relation based on
672 stars has a slope of $-3.193 \pm 0.027$. 
Using the Monte Carlo
method with samples of 66 stars the mean slope is $-3.23 \pm 0.05$.

The conclusion is that, within the errors, the distribution of period and/or a
non-linearity effect does not bias the derived slope in $V$ and $K$.

In $V$ the observed slope in the LMC agrees within
1$\sigma$ with the one derived for the Galactic sample both for a
constant $p$-factor and for the one with a mild period dependence. 
In $K$ the agreement is less good, in the sense that
the $p$-factor with a dependence on period agrees more closely.

\section{Summary and discussion} 

The period-luminosity in the $V$ and $K$-band and the dependence on
metallicity are investigated for a sample of 68 Galactic Cepheids with
an individual metallicity determination from high-resolution spectroscopy. 
Distances are derived using the Baade-Wesselink technique using the
most recent surface-brightness relation and estimates for the projection factor.

The difference in slope between a constant $p$-factor (consistent with
available interferometric data) and a mild dependence on period
(following theoretical work by Nardetto et al. 2007) results in a
difference in slope in the $PL$-relation which is of the same
magnitude as the error in slope.

The slope of the LMC $PL$-relation in $V$ and $K$ is derived taking
into account the difference in period distribution between the
Galactic sample and the LMC sample. In $V$ the slope for the LMC
sample agrees with the galactic one for both assumptions regarding the
dependence of $p$ on period. In $K$ the comparison of the slopes
favours the mild period dependence of the $p$-factor on period.

The metallicity dependence of the $PL$-relation is investigated and no
significant dependence is found. A firm result is not possible as the
range in metallicity spanned by the current sample of galactic
Cepheids is 0.3-0.4 dex, while previous work suggested a small
dependence on metallicity only (typically $-0.2$ mag/dex, see the introduction). 

It would be extremely valuable to analyse a sample of LMC and SMC
Cepheids in a similar way. Unfortunately, the sample of Cepheids with
adequate optical, NIR and RV light curves (see e.g.  Storm et
al. 2004, 2005) for which a Baade-Wesselink distance has been derived
(Gieren et al. 2005) and the sample of Magellanic Cloud Cepheids for
which a metallicity has been derived from high-resolution spectroscopy
(Mottini et al. 2008) have no overlap. High-resolution spectroscopy of
MC Cepheids with Baade-Wesselink distances would therefore be
extremely valuable.

\acknowledgements{  
I would like to thank Dr. Igor Soszy\'nski for communicating the intensity-mean 2MASS magnitudes for the OGLE Cepheids, and
Maryline Briquet and Djazia Ladjal for taking the Coralie data. 
The referee pointed out additional metallicity determinations in the literature which led to a significantly larger sample of stars under study.
I also thank the ESO librarians Uta Grothkopf and Chris Erdmann for obtaining 
some of the century old papers with radial velocity data.
Part of this research was done while MG was a short-term visitor at the Max-Planck Institut f\"ur Astrophysik (MPA) in Garching, Germany.
This research has made use of the SIMBAD database, operated at CDS, Strasbourg, France. 
}

{}


\begin{thebibliography}{} 

\bibitem[]{} Abt, H.A. 1959, AJ, 130, 1021
\bibitem[]{} Abt, H.A. 1973, ApJS, 26, 365
\bibitem[]{} Abt, H.A., \&  Levy, S.G. 1978, PASP, 90, 188
\bibitem[]{} Adams, W.S., \& Shapley, H. 1918 ApJ, 47, 46
\bibitem[]{} Albrow, M.D., \& Cottrell, P.L. 1996, MNRAS, 280, 917
\bibitem[]{} Alcock, C., Allsman, R.A., Axelrod, T.S., Bennett, D.P., Cook, K.H., 1995, AJ, 109, 1653
\bibitem[]{} Andrievsky, S.M., Bersier, D., Kovtyukh, V.V., et al. 2002b, A\&A, 384, 140
\bibitem[]{} Andrievsky, S.M., Egorova, I.A., Korotin, S.A., \& Kovtyukh, V.V. 2003, AN, 324, 532 
\bibitem[]{} Andrievsky, S.M., Kovtyukh, V.V., Luck, R.E., et al. 2002a, A\&A, 381, 32
\bibitem[]{} Andrievsky, S.M., Kovtyukh, V.V., Luck, R.E., et al. 2002c, A\&A, 392, 491
\bibitem[]{} Andrievsky, S.M., Luck, R.E., \& Kovtyukh, V.V. 2005, AJ, 130, 1880
\bibitem[]{} Andrievsky, S.M., Luck, R.E., Martin P., \& L\'epine, J.R.D. 2004, A\&A, 413, 159
\bibitem[]{} Babel, J., Burki, G., Mayor, M., Chmielewski, Y., Waelkens, C. 1989, A\&A, 216, 125
\bibitem[]{} Barnes, T.G., Fernley, J.A., Frueh, M.L., Navas, J.G., Moffett, T.J., \& Skillen, I. 1997, PASP, 109, 645
\bibitem[]{} Barnes, T.G., Jeffery, E.J., Berger, J.O., Mueller, P.J., Orr, K., \& Rodriguez, R. 2003, ApJ, 592, 539
\bibitem[]{} Barnes, T.G., Jeffery, E.J., Montemayor, T.J., Skillen, I. 2005, ApJS, 156, 227
\bibitem[]{} Barnes, T.G., Moffett, T.J., \& Slovak, M.H. 1987, ApJS, 65, 307
\bibitem[]{} Barnes, T.G., Moffett, T.J., \& Slovak, M.H. 1988, ApJS, 66, 43
\bibitem[]{} Barnes, T.G., Storm, J., Jefferys, W.J., Gieren, W.P., \& Fouqu\'e, P. 2005, ApJ, 631, 572
\bibitem[]{} Beavers, W.I. \& Eitter, J.J. 1986, ApJS, 62, 147
\bibitem[]{} Benedict, G.F., McArthur, B.E., Feast, M.W., Barnes, T.G., Harrison, T.E., et al. 2007, AJ, 133, 1810 
\bibitem[]{} Berdnikov, L.N., Dambis, A.K.,\& Vozyakovs, O.V. 2000, A\&AS, 143, 211
\bibitem[]{} Bersier, D. 2002, ApJS, 140, 465
\bibitem[]{} Bersier, D., Burki, G., Mayor, M., \& Duquennoy, A. 1994, A\&AS, 108, 25
\bibitem[]{} B\"ohm-Vitense, E., Clark, M., Cottrell, P.L., \& Wallerstein, G. 1990, AJ, 99, 353
\bibitem[]{} Breger, M. 1967 MNRAS, 136, 61
\bibitem[]{} Breger, M. 1970, AJ, 75, 239
\bibitem[]{} Breitfellner, M.G., \& Gillet, D. 1993, A\&A, 277, 541
\bibitem[]{} Caldwell, J.A.R., Coulson, I.M., Dean, J.F., \& Berdnikov, L.N. 2001, JAR 7, 4
\bibitem[]{} Campbell, W.W., \& Moore, J.H. 1928  Lick Obs. Bull., 16, 267
\bibitem[]{} Carter, B.S. 1990, MNRAS, 242, 1
\bibitem[]{} Coulson, I.M. 1983, MNRAS, 203, 925
\bibitem[]{} Coulson, I.M., \& Caldwell, J.A.R. 1985a, SAAO Circulars, 9, 5
\bibitem[]{} Coulson, I.M., \& Caldwell, J.A.R. 1985b, MNRAS, 216, 671
\bibitem[]{} Coulson, I.M., Caldwell, J.A.R., \& Gieren, W.P. 1985, ApJS, 57, 595
\bibitem[]{} Dean, J.F. 1977, MNSSA, 36, 3
\bibitem[]{} Dean, J.F. 1981, SAAOC, 6, 10
\bibitem[]{} Duncan, J.C. 1908, Lick Obs. Bull., 5, 82
\bibitem[]{} Duncan, J.C. 1922, ApJ, 56, 340
\bibitem[]{} Duncan, J.C. 1932, PASP 44, 324
\bibitem[]{} Eggen, O.J. 1983, AJ, 88, 379
\bibitem[]{} Evans, N.R. 1976, ApJS, 32, 399
\bibitem[]{} Evans, N.R., 1988, ApJS, 66, 343  
\bibitem[]{} Evans, N.R., Carpenter, K., Robinson, R., Massa, D., Wahlgren, G.M., Vink\'o, J., \& Szabados, L. 1999, ApJ, 524, 379 
\bibitem[]{} Evans, N.R., \& Lyons, R. 1986, AJ, 92, 436
\bibitem[]{} Evans, N.R., \& Lyons, R. 1994, AJ, 107, 2164
\bibitem[]{} Evans, N.R., \& Sugars, B.J.A. 1997, AJ, 113, 792
\bibitem[]{} Evans, N.R., \& Welch, D.L. 1993, PASP, 105, 836
\bibitem[]{} Evans, N.R., Welch, D.L., Scarfe, C.D., \&  Teays, T.J. 1990, AJ, 99, 1598  
\bibitem[]{} Evans, N.R., Welch, D.L., Slovak, M.H., Barnes, T.G. III, \& Moffett, T.J. 1993, AJ, 106, 1599
\bibitem[]{} Feast, M.W. 1967, MNRAS, 136, 141
\bibitem[]{} Feast, M.W. 1999, PASP, 111, 775
\bibitem[]{} Feast, M.W., Laney, C.D., Kinman, T.D., van Leeuwen, F., \& Whitelock, P.A. 2008, MNRAS, 386, 2115  
\bibitem[]{} Fernley, J.A., Skillen I., \& Jameson, R.F. 1989, MNRAS 237, 947
\bibitem[]{} Fouqu\'e, P., Arriagada, P., Storm, J., Barnes, T.G., Nardetto N., et al. 2007, A\&A, 476, 73  (F07)
\bibitem[]{} Fouqu\'e, P., \& Gieren, W.P. 1997, A\&A, 320, 799
\bibitem[]{} Fouqu\'e, P., Storm, J., \& Gieren, W.P. 2003, in: ``Stellar candles for the extragalactic distance scale'', Lect. Notes Phys., 635, 21
\bibitem[]{} Frost, E.B. 1906, ApJ, 23, 264
\bibitem[]{} Fry, A.M., Carney, B.W., 1997, AJ, 113, 1073
\bibitem[]{} Gieren, W.P. 1976, A\&A 47, 211
\bibitem[]{} Gieren, W.P. 1981a, ApJS, 46, 287
\bibitem[]{} Gieren, W.P. 1981b, ApJS, 47, 315
\bibitem[]{} Gieren, W.P. 1985, ApJ, 295, 507
\bibitem[]{} Gieren, W.P. 1989a, A\&A, 216, 135
\bibitem[]{} Gieren, W.P. 1989b, PASP, 101, 160
\bibitem[]{} Gieren, W.P., Barnes T.G., \& Moffett, T.J., 1993, ApJ 418, 135
\bibitem[]{} Gieren, W.P., Fouqu\'e, P., G\'omez, M. 1997, ApJ 488, 74
\bibitem[]{} Gieren, W.P., Fouqu\'e, P., G\'omez, M. 1998, ApJ 496, 17
\bibitem[]{} Gieren, W.P., Storm, J., Barnes, T.G., Fouqu\'e, P., Pietrzy\'nski, G., \& Kienzle, F. 2005, ApJ, 627, 224
\bibitem[]{} Glass, I.S. 1985, IrAJ, 17, 1
\bibitem[]{} Gorynya, N.A., Samus, N.N., Berdnikov, L.N., Rastorgouev, A.S., \& Sachkov, M.E., 1995, IBVS, 4199
\bibitem[]{} Gorynya, N.A., Samus', N.N., Sachkov, M.E., Rastorguev, A.S., Glushkova, E.V., \& Antipin, S.V. 1998, PAZh, 24, 939 (VizieR On-line Data Catalog: III/229)
\bibitem[]{} Gould, A., 1994, ApJ, 426, 542
\bibitem[]{} Grayzeck, E.J. 1978, AJ, 83, 1397
\bibitem[]{} Groenewegen, M.A.T. 2007, A\&A, 474, 975
\bibitem[]{} Groenewegen, M.A.T., Romaniello, M., Primas, F., \& Mottini, M. 2004, A\&A, 420, 655
\bibitem[]{} Harper, W.E. 1934, Publ. DAO, 6, 151
\bibitem[]{} H\"aupl, W. 1988, AN, 309, 3
\bibitem[]{} Henden, A.A. 1980, MNRAS, 192, 621
\bibitem[]{} Herbig, G.H., \& Moore, J.H. 1952, ApJ, 116, 348
\bibitem[]{} Imbert, M. 1985, A\&AS, 58, 529
\bibitem[]{} Imbert, M. 1996, A\&AS, 116, 497   
\bibitem[]{} Imbert, M. 1999, A\&AS, 140, 79
\bibitem[]{} Isobe, T., Feigelson, E.D., Akritas, M.G., \& Babu, G.J. 1990, ApJ, 364, 104
\bibitem[]{} Jacobsen, T.S. 1970 ApJ, 159, 569
\bibitem[]{} Jacobsen, T.S. 1974, ApJ, 191, 691
\bibitem[]{} Jacobsen, T.S., \& Wallerstein, G. 1981, PASP, 93, 481
\bibitem[]{} Joy, H.A. 1937, ApJ, 86, 363
\bibitem[]{} Joy, H.A. 1952, ApJ, 115, 25
\bibitem[]{} Kennicutt, R.C., Stetson, P.B., Saha, A., et al., 1998, ApJ, 498, 181
\bibitem[]{} Kervella, P., Bersier, D., Mourard D.,  Nardetto, N., Fouqu\'e, P., \& Coud\'e du Foresto, V. 2004b, A\&A, 428, 587 
\bibitem[]{} Kimeswenger,S., Lederle, C., Richichi, A. et al. 2004, A\&A, 413, 1037
\bibitem[]{} Kiss, L.L. 1998, MNRAS, 297, 825
\bibitem[]{} Kiss, L.L. 2000, MNRAS, 314, 420
\bibitem[]{} Kochanek, C.S., 1997, ApJ, 491, 13
\bibitem[]{} Koen, C., Marang, F., Kilkenny, D., \& Jacobs, C. 2007, MNRAS, 380, 1433
\bibitem[]{} Kovtyukh V.V., Andrievsky, S.M, Belick, S.I., \& Luck, R.E. 2005b, AJ, 129, 433
\bibitem[]{} Kovtyukh V.V., Wallerstein, G. \& Andrievsky, S.M, 2005a, PASP, 117, 1173
\bibitem[]{} Laney, C.D. \& Caldwell, J.A.R. 2007, MNRAS, 377, 147
\bibitem[]{} Laney, C.D. \& Stobie, R.S. 1992, A\&AS, 93, 93
\bibitem[]{} Lemasle, B., Francois, P., Bono, G., Mottini, M., Primas, F., \& Romaniello M. 2007, A\&A, 467, 283
\bibitem[]{} Lenz, P., \& Breger, M. 2005, CoAst, 146, 53
\bibitem[]{} Lloyd Evans, T. 1968, MNRAS, 141, 109
\bibitem[]{} Lloyd Evans, T. 1980, SAAO Circulars, 1, 163
\bibitem[]{} Lloyd Evans, T. 1980, SAAO Circulars, 1, 257
\bibitem[]{} Luck, R.E., \& Andrievsky, S.M. 2004, AJ, 128, 343
\bibitem[]{} Luck, R.E.,  Kovtyukh V.V., \& Andrievsky, S.M. 2006, AJ, 132, 902
\bibitem[]{} Luck, R.E., Gieren, W.P., Andrievsky, S.M., et al. 2003, A\&A, 401, 939
\bibitem[]{} Macri, L.M., Stanek, K.Z., Bersier, D., Greenhill, L.J., Reid, M.J. 2006, ApJ, 652, 1113
\bibitem[]{} Maddrill, J.D. 1906, PASP, 18, 252
\bibitem[]{} Madore, B.F. 1975, ApJS, 29, 219
\bibitem[]{} M\'erand, A., Kervella, P., Coud\'e du Foresto, V. et al. 2005, A\&A, 438, L9
\bibitem[]{} Mermilliod, J.-C., Mayor, M., \& Burki, G. 1987 A\&AS, 70, 389
\bibitem[]{} Metzger, M.R., Caldwell, J.A.R., McCarthy, J.K., \& Schechter, P.L. 1993, ApJS, 76, 803
\bibitem[]{} Metzger, M.R., Caldwell, J.A.R., \& Schechter, P.L. 1992, AJ, 103, 529
\bibitem[]{} Moffett, T.J., \& Barnes, T.G. 1984, ApJS, 55, 389
\bibitem[]{} Moore, J.H. 1909, Lick Obs. Bull., 5, 111
\bibitem[]{} Mottini, M., Romaniello, M., Primas, F., Pedicelli, S., Lemasle, B., Bono, G., 
Fran\c{c}ois, P., Groenewegen, M.A.T., \&  Laney, C.D. 2008, A\&A, submitted  
\bibitem[]{} Nardetto, N., Fokin, A., Mourard, D., Mathias, Ph., Kervella, P., \& Bersier, D. 2004, A\&A, 428, 131
\bibitem[]{} Nardetto, N., Mourard, D., Kervella, P., Mathias, Ph., M\'erand, A., \& Bersier, D. 2006, A\&A, 453, 309
\bibitem[]{} Nardetto, N., Mourard, D., Mathias, Ph., Fokin, A., \& Gillet, D., 2007, A\&A, 471, 661
\bibitem[]{} Ngeow, C., Kanbur, S.M., Nanthakumar, A. 2008, A\&A, 477, 621
\bibitem[]{} Niva, G.D., \& Schmidt, E.G. 1979, ApJ, 234, 245
\bibitem[]{} Pel, J.W. 1976, A\&AS, 24, 413
\bibitem[]{} Persson, S.E., Madore, B.F., Krzeminski, W, et al. 2004, AJ, 128, 2239
\bibitem[]{} Petterson, O.K.L., Cottrell, P.L. \& Albrow, M.D., 2004, MNRAS, 350, 95 
\bibitem[]{} Petterson, O.K.L., Cottrell, P.L., Albrow, M.D., \& Fokin, A. 2005, MNRAS 362, 1167
\bibitem[]{} Pont, F., Burki, G., \& Mayor, M. 1994, A\&AS, 105, 165
\bibitem[]{} Press, W.H., Teukolsky, S.A., Vetterling, W.T.,\& Flannery, B.P. 1992, in Numerical Recipes in Fortran 77, Cambridge U.P.
\bibitem[]{} Rastorgouev, A.S., Gorynya, N.A., \& Samus, N.N., 1997, in Binary Stellar Systems,  ed. A.G. Massevich (Moscow, Kozmosinform), p. 123
\bibitem[]{} Sanford, R.F. 1927, ApJ, 66, 170
\bibitem[]{} Sanford, R.F. 1935, ApJ, 81, 140
\bibitem[]{} Sanford, R.F., 1951, ApJ, 114, 331
\bibitem[]{} Sanford, R.F., 1956, ApJ, 123, 201
\bibitem[]{} Sasselov, D.D., Beauliau, J.P., Renault, C.,  et al., 1997, A\&A, 324, 471
\bibitem[]{} Schechter, P.L., Avruch, I.M., Caldwell, J.A.R., \& Keane, M.J. 1992, AJ, 104, 1930
\bibitem[]{} Schwarz, G., 1978, Ann. Stat., 6, 461
\bibitem[]{} Shobbrook R.R. 1992, MNRAS, 255, 486
\bibitem[]{} Slipher, V.M. 1904, ApJ, 20, 146
\bibitem[]{} Soszy\'nski, I., Gieren, W., Pietrzy\'nski, G., et al. 2005, PASP, 117, 823
\bibitem[]{} Stibbs, D.W.N. 1955, MNRAS, 115, 363
\bibitem[]{} Storm, J., Carney, B.W., Gieren, W.P., et al., 2004, A\&A, 415, 531
\bibitem[]{} Storm, J., Carney, B.W., Gieren, W.P., Fouqu\'e, P., Latham, D.W., \& Fry, A.M. 2004, A\&A, 415, 531
\bibitem[]{} Storm, J., Gieren, W.P., Fouqu\'e, Barnes, T.G., \& G\'omez, M. 2005, A\&A, 440, 487
\bibitem[]{} Sugars, B.J.A., \& Evans, N.R. 1996, AJ, 112, 1670
\bibitem[]{} Szabados, L. 1977, Mitt. Sternw. Ung. Akad. Wiss., Budapest, No.70
\bibitem[]{} Szabados, L. 1980, Commun. Konkoly Obs. Hung. Acad. Sci., Budapest, No.76
\bibitem[]{} Szabados, L. 1981, Commun. Konkoly Obs. Hung. Acad. Sci., Budapest, No.77
\bibitem[]{} Szabados, L. 1989, CoKon, 94, 1
\bibitem[]{} Szabados, L. 1990, MNRAS, 242, 285 
\bibitem[]{} Szabados, L. 1991, Commun. Konkoly Obs. Hung. Acad. Sci., Budapest, No.96
\bibitem[]{} Szabados, L. 1996, A\&A, 311, 189
\bibitem[]{} Szabados, L. 2003, Inf. Bull. Var. Stars, No. 5394
\bibitem[]{} Taylor, M.M.,  Albrow, M.D., Booth, A.J., Cottrell, P.L. 1997 MNRAS, 292, 662
\bibitem[]{} Taylor, M.M., \& Booth, A.J. 1998, MNRAS, 298, 594
\bibitem[]{} ten Bruggencate, P. 1930, Harvard Circ., no 351, 1
\bibitem[]{} Turner, D.G., Bryukhanov, I.S., Balyuk, I.I., Gain, A.M., Grabovsky, R.A., et al. astro-ph/0709.3085
\bibitem[]{} Udalski, A., Soszy\'nski, I., Szyma\'nski, M., et al., 1999, AcA, 49, 223 
\bibitem[]{} van Belle, G.T., Lane, B.F., \& Thompson, R.R. 1999, AJ, 117, 521
\bibitem[]{} van Leeuwen, F. 2007, ``Hipparcos, the new reduction of the Raw Data'', ASSL 350, Springer
\bibitem[]{} van Leeuwen, F., Feast, M.W., Whitelock, P.A., \& Laney, C.D. 2007, MNRAS, 379, 23
\bibitem[]{} Wallerstein, G. 1972, PASP, 84, 656
\bibitem[]{} Walraven, J.H., Tinbergen, J., \& Walraven, T. 1964, BAN, 17, No. 7, 520
\bibitem[]{} Welch, D.L. 1985, PhD thesis, University of Toronto
\bibitem[]{} Welch, D.L., Wieland, F., McAlary, C.W., et al. 1984, ApJS, 54, 547
\bibitem[]{} Wilson, T.D., Carter, M.W., Barnes, T.G., Van Citters, G.W., \& Moffett, T.J. 1989, ApJS, 69, 951
\bibitem[]{} Wisniewski, W.Z., \& Johnson, H.R. 1968, CoLPL, 7, 57
\bibitem[]{} Yong, D., Carney, B.W., Teixera de Almeida, M.-L., \& Pohl, B.L. 2006, AJ, 131, 2256

\end{thebibliography}
\end{document}